\documentclass{aastex701}

\usepackage{amssymb}
\usepackage{amsmath}
\usepackage{graphicx}
\usepackage{booktabs}
\usepackage{multirow}
\usepackage{siunitx}
\usepackage{url}
\usepackage{aas_macros}
\usepackage{natbib}
\usepackage{threeparttable}
\usepackage{footmisc}

\begin{document}

\title{Is the `Known' Enough? An Integrated Machine Learning Framework for Eclipsing Binary Classification and Parameter Estimation Based on Well-Characterized Systems}

\author[orcid=0000-0002-4624-3847,sname='Ula\c{s}']{Burak Ula\c{s}}
\affiliation{Department of Space Sciences and Technologies, Faculty of Sciences,
Çanakkale Onsekiz Mart University, Terzioğlu Campus,
Çanakkale TR-17100, Turkey}
\email[show]{burak.ulas@comu.edu.tr}  








\begin{abstract}

This study presents a multi-task machine learning framework for simultaneous morphology classification and physical parameter estimation of eclipsing binaries using photometric light curves. We train Random Forest and XGBoost ensemble models on 845 of 995 well-characterized systems comprising three morphological configurations by extracting 51 domain-specific features from each phase-folded light. To assess generalization, 15\% of systems were withheld as an independent test set before any model training. On this held-out set, the XGBoost model yields $R^2$ values of 0.88 for the effective temperature ratio, 0.91 for the primary surface potential, 0.92 for the secondary surface potential, 0.89 for inclination, and 0.77 for the mass ratio. Morphology classification achieves 95.4\% accuracy on the cross-validation set with per-class F1 scores exceeding 0.90, while the held-out test set confirms generalization with 90.7\% accuracy. To demonstrate the model's scalability, we applied the trained framework to over 100000 uncharacterized light curves from the OGLE and {\it{Kepler}} eclipsing binary databases, enforcing strict physics-based constraints on the Roche potentials during prediction to ensure astrophysical validity. We present a catalog of estimated physical parameters and classifications for these systems, identifying thousands of high-confidence candidates. Morphological classifications are independently validated against the OGLE Online Catalog of Variable Stars (OCVS), achieving a contact recall of 0.99 across 104692 matched systems. The model's generalization capability is validated by cross-matching predictions with independent Kepler catalogs, achieving 77\%  classification accuracy and recovering physical parameters with systematic deviations consistent with known selection biases, third-light dilution, and methodological differences between photometric and spectroscopic approaches. This work confirms that machine learning ensembles, when coupled with physics guided post-processing, can effectively bridge the gap between massive photometric surveys and detailed astrophysical characterization.

\end{abstract}



\section{Introduction}
\label{sec:intro}

Eclipsing binary stars constitute a foundational pillar of modern astrophysics, serving as the primary empirical source for determining fundamental stellar parameters such as mass, radius, and luminosity. These systems offer unique constraints on stellar evolution models and the dynamics of interacting systems. Historically, the analysis of these binaries has relied on conventional modeling codes, which employ iterative differential correction or simplex minimization to solve the inverse problem. Although these traditional frameworks yield robust physical parameters, they are computationally intensive and require significant manual intervention to resolve geometric degeneracies, making them ill-suited for the rapid characterization of large statistical samples.

The overwhelming number of photometric data from ground- and space-based surveys has necessitated a paradigm shift from traditional analysis to automated computational techniques. The early foundational work by \citet{Sar06} and \citet{Arm16} demonstrated the viability of the ensemble methods, employing Bayesian neural network ensembles and hybrid Random Forest techniques with Self-Organizing Maps, respectively, for the morphological classification of eclipsing binaries. Building on this, \citet{Che20} addressed the challenges of ground-based surveys by processing ZTF Data Release 2 using an auto-manual DBSCAN clustering method based on parameters such as period, Fourier coefficients, amplitude and absolute Wesenheit magnitude. Their aim was to triage millions of variable sources, successfully identifying approximately 350000 eclipsing binaries with high accuracy. Similarly, for the Gaia mission, \citet{Mow23} implemented a multi-stage pipeline that utilized geometric modeling of G-band light curves to handle sparse sampling. This effort resulted in the Gaia DR3 catalog of more than 2.1 million eclipsing binary candidates, demonstrating that automated pipelines could effectively filter and classify sources even in datasets with irregular observing cadences.

In recent years, the field has evolved from simple feature-based classification to deep learning architectures capable of direct physical characterization. \citet{Parimucha2024} introduced a computer vision approach, transforming phase-folded light curves into 2D polar images and analyzing them with Convolutional Neural Networks (CNNs) like ResNet50 to capture morphological distinctions without manual feature extraction. To address the complexities of high-cadence data, \citet{Shan2025} developed a hybrid architecture that combines CNNs, LSTMs, and attention mechanisms to extract both local eclipse features and global temporal dependencies, thereby identifying thousands of new candidates in the TESS data. Notably, recent studies have tackled the "inverse problem" of parameter estimation. \citep{Ding2024} pioneered the use of Autoencoder neural networks trained on synthetic PHOEBE models to detect contact binary candidates. Extending this to massive catalogs, \citet{Li2025} employed fully connected neural networks to derive physical parameters ($q,i,T_{e_2}/T_{e_1}$) for over 12000 contact binaries, crucially incorporating spot parameters to model the O'Connell effect. Furthermore, \citet{Xiong2024} applied a similar forward-modeling pipeline to characterize semidetached systems. However, as contemporary machine learning models require large amounts of data for training, we still face the lack of a comprehensive light curve parameter and data catalog for binary stars, especially from space missions.

A central question motivating this work is whether a modest but well-characterized training sample, which comprises fewer than 1000 systems with physical parameters derived from detailed multi-band light curve solutions, is sufficient to train machine learning models that generalize to the broader eclipsing binary population.  We hypothesize that such a sample is sufficient, provided the training set spans the primary morphological regimes (detached, semidetached, and contact) and the physically relevant parameter space. To test this hypothesis, we quantify the separability of the three morphological classes through pairwise Kolmogorov--Smirnov tests, validate the trained models on strictly held-out catalogue systems, apply the  framework to large-scale survey data, and examine the feature-space coverage overlap between the training set and the target populations.

The remainder of this paper is organized as follows. Sec.~\ref{sec:data} describes the curation of our multi-source training set of 995 well-characterized eclipsing binaries and the extraction of 51 domain-specific features from phase-folded light curves. Sec.~\ref{sec:training} presents the Random Forest \citep[RF;][]{bre01} and XGBoost \citep[XGB;][]{che16} ensemble architectures, together with hyperparameter sensitivity,  multi-output, and overfitting analyses. In Sec.~\ref{sec:predictions}, we apply the trained framework to over 100000 light curves from the OGLE and \textit{Kepler} eclipsing binary databases, validate the predictions against two catalogs, and demonstrate the pipeline's scalability. Finally, Sec.~\ref{sec:discussion} discusses the astrophysical implications of the results, confirms through a Principle Component Analysis (PCA) coverage analysis that the training sample spans the high-density regions of both survey populations, and addresses the role of observational selection effects and the computational advantages of the framework for future surveys.

\section{Data}
\label{sec:data}

\subsection{Dataset Description and Preprocessing}
\label{sec:dataset}
\begin{sloppypar}

Our training sample comprises 995 eclipsing binary systems compiled from three primary sources. The fundamental source is the Catalog and Atlas of Eclipsing Binaries \citep[CALEB\footnote{\url{http://caleb.eastern.edu}}, former EBOLA,][]{bra04}, from which we extracted all available photometric bands for each system. The geometric and physical parameters for these systems were originally derived through formal light curve solutions using the Binary Maker 3 code, incorporating multi-band photometry and, where available, radial velocity curves to constrain the system geometry. Within the CALEB archive, systems explicitly classified as 'overcontact' were merged into the 'contact' category. This aggregation maintains a statistically significant sample size for the contact class and is methodologically justified, as overcontact binaries share the essential morphological signatures (e.g. continuous flux variations, equal minima depths) required for robust classifier training. Two supporting catalogs supplemented this collection: \textsc{DEBC}at\footnote{\url{https://www.astro.keele.ac.uk/jkt/debcat}} \citep{sou15}, a catalog of detached eclipsing binary stars, and {\textsc{W UM}}a {\textsc{C}}at\footnote{\url{https://wumacat.aob.rs}}, the compilation of 700 individually studied W UMa stars by \citet{lat21}. For these two supporting sources, systems were included only if their light curve data were publicly available through the Centre de Donn\'{e}es astronomiques de Strasbourg (CDS). For systems in the  \textsc{DEBC}at and {\textsc{W UM}}a {\textsc{C}}at catalogs where pre-calculated orbital phases were not provided, we computed the phases using the orbital periods and times of minimum light calculated from the data given by the source. Additionally, to ensure a consistent input format for feature extraction, any photometric data provided in magnitudes were converted to flux units prior to processing.

For training data quality and physical consistency, systems were excluded from the initial compilation if they met any of the following criteria: (1) large gaps in orbital phase coverage preventing reliable eclipse characterization; (2) extreme photometric scatter indicating poor data quality; (3) light curve morphology inconsistent with eclipsing binary signatures; or (4) absence of both primary and secondary minima. Furthermore, every light curve in the training set underwent a visual inspection to guarantee the absence of artifacts or non-physical outliers that could compromise the normalization process. For the training dataset, comprised of high-quality, visually verified light curves, we employed standard normalization by dividing by the absolute maximum flux. However, when applying the model to large-scale surveys like OGLE, we switched to a robust 99.5th percentile normalization. This ensures that instrumental artifacts or cosmic ray outliers do not distort the scaling, effectively matching the noisy observational data to the clean dynamic range of the training set (see Sec.~\ref{sec:predictions}).

\begin{figure*}
\centering
\plotone{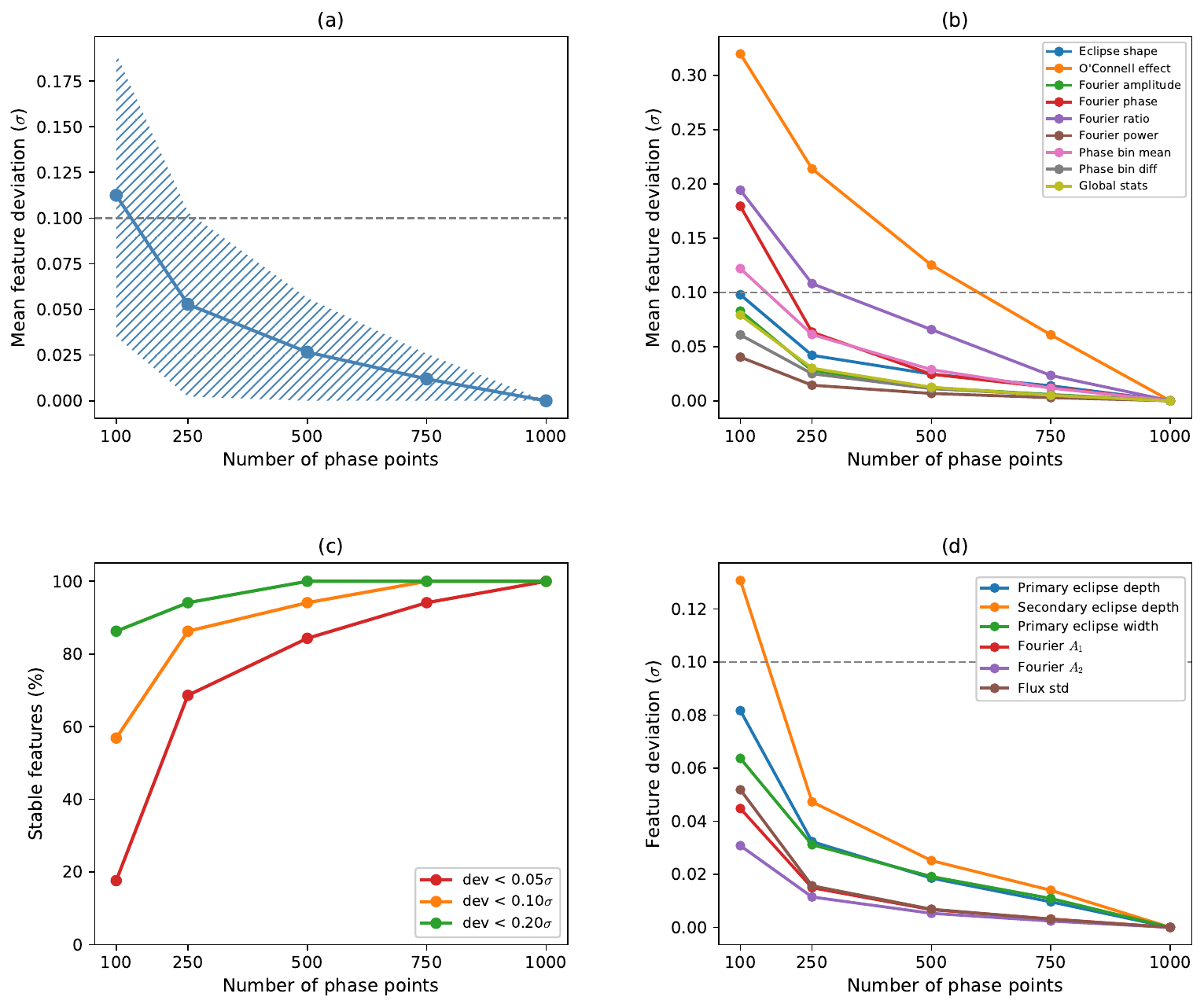}
\caption{Resolution sensitivity of the 51 extracted features. Each panel shows results for subsampling at 100, 250, 500, and 750 phase points (the baseline is 1000 points). Feature deviation is measured in units of the training-set standard deviation ($\sigma$) of each feature. The horizontal dashed line marks the 10\% threshold ($0.10\sigma$) used throughout the analysis. (a) Solid line: overall mean deviation across all 51 features and all 995 light curves; shaded band corresponds to $\pm1\sigma$ spread across the 51 per-feature mean deviations. (b) Per-group mean deviation for nine feature groups: \textit{Eclipse shape} (primary/secondary eclipse depth and width; 4 features), \textit{O'Connell effect} (3 out-of-eclipse asymmetry features), four Fourier sub-groups (amplitudes, phases, ratios, and total power; 17 features), two phase-bin sub-groups (bin means and consecutive differences; 19 features), and \textit{Global stats} (8 basic flux statistics; individual feature definitions are available at \url{https://github.com/burakulas/ml-ebs/blob/main/Feature_List.md}). (c) Fraction of features with deviation below $0.05\sigma$, $0.10\sigma$, and $0.20\sigma$. (d) Deviation of six representative features spanning different feature types.} 
\label{fig:resolution}
\end{figure*}

All light curves were phase-folded to the interval $[0.25, 1.25]$ with the primary (deeper) eclipse centered near phase 1.0. Sparse observational data were interpolated to 1000 uniformly spaced phase points using PCHIP \citep[Piecewise Cubic Hermite Interpolating Polynomial;][]{fri80} interpolation, which preserves monotonicity and avoids oscillations near eclipse minima. A resolution sensitivity analysis was performed to characterize the minimum sampling required for reliable feature extraction. For each of the 995 training light curves we generated subsampled versions at 100, 250, 500, and 750 phase points by randomly drawing the corresponding number of points from the full 1000-point grid, re-applying PCHIP interpolation to the standard grid, and extracting all 51 features. We repeated this procedure ten times per light curve per resolution level and measured the per-feature deviation from the full-resolution baseline in units of the training-set standard deviation of each feature. Fig.~\ref{fig:resolution}a shows the overall mean deviation averaged across all 51 features as a function of sampling density. The results show that the mean deviation across all 51 features falls below 0.1$\sigma$ at 750 phase points, with all but the O'Connell effect features individually below this threshold (Fig.~\ref{fig:resolution}b) and 94\% of features are stable already at 500 points (Fig.~\ref{fig:resolution}c). The 1000-point grid is therefore conservative. The features most sensitive to undersampling are those measuring the out-of-eclipse flux asymmetry (deviation is about $0.32\sigma$ at 100 points) and the eclipse-wing phase bins, while Fourier amplitudes and global statistics remain stable even at 100 points (deviation $<0.06\sigma$, Fig.~\ref{fig:resolution}b,d). We therefore recommend applying this framework to light curves with at least 500 phase-folded observations, as is typical of OGLE, \textit{Kepler}, and TESS time-series data. When it comes to interpolation, PCHIP is monotonicity-preserving and introduces no oscillatory artefacts at eclipse boundaries. The sensitivity analysis demonstrates that features extracted after PCHIP reconstruction converge rapidly with sampling density, confirming that the interpolation does not introduce systematic bias above the $0.1\sigma$ level for light curves with 500 or more observations. 

We adopted a mass ratio definition of $q = M_2/M_1$ with a tolerance of $q \leq 1.05$, allowing for small measurement uncertainties near unity. Systems in the input catalog with $q > 1.05$ underwent a mathematically equivalent transformation: $q_{\text{new}} = 1/q$. This inversion included the simultaneous swapping of component-specific parameters like effective temperatures ($T_{e_1}, T_{e_2}$) and Roche surface potentials ($\Omega_1, \Omega_2$), as well as a phase shift of 0.5 to preserve eclipse alignment. After this transformation, systems with $T_{e_1}, T_{e_2} > 1.0$ remain in the training set (Table~\ref{tab:parameters}) and these correspond to semidetached Algol-type binaries in which mass transfer has reversed the mass-luminosity relation, leaving the currently less massive component as the hotter star.

\begin{figure*}
	\centering
	\plotone{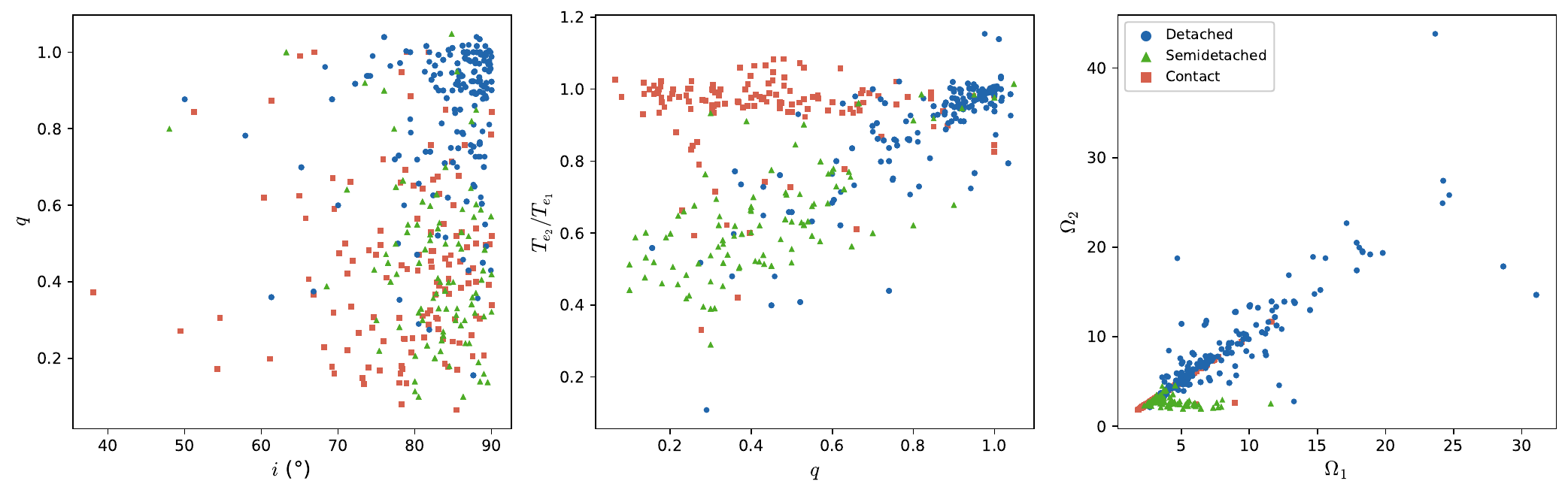}
	\caption{Parameter space of the 995 training systems ($N_\mathrm{Detached}=421$, $N_\mathrm{Semidetached}=234$, $N_\mathrm{Contact}=340$) shown in three projection planes: inclination vs.\ mass ratio ($i$, $q$; left), mass ratio vs.\ temperature ratio ($q$, $T_{e_2}/T_{e_1}$; middle), and primary vs.\ secondary surface potential ($\Omega_1$, $\Omega_2$; right).}
	\label{fig:param_space}
\end{figure*}

The final curated sample contains three morphological classes: 421 detached systems (42.3\%), 340 contact binaries (34.2\%), and 234 semidetached configurations (23.5\%). This relatively balanced distribution enables effective multi-class classification without requiring class weighting or oversampling techniques. Table~\ref{tab:parameters} summarizes the parameter distributions in our training sample. Notably, the training set exhibits a strong observational bias toward high inclinations ($i = 82.0^{\circ} \pm 7.6^{\circ}$). This reflects the unavoidable selection effect where edge-on systems produce deeper, more detectable eclipses. Figure~\ref{fig:param_space} also shows the distribution of the 995 training systems in three parameter-space projections: inclination versus mass ratio ($i$, $q$), mass ratio versus temperature ratio ($q$, $T_{e_2}/T_{e_1}$), and primary versus secondary surface potential ($\Omega_1$, $\Omega_2$). The three morphology classes occupy distinct regions of this space. In the ($q$, $T_{e_2}/T_{e_1}$) plane the separation is most pronounced: detached and contact systems concentrate near $T_{e_2}/T_{e_1} \approx 0.90$--$0.95$, while semidetached systems occupy a lower temperature-ratio band ($T_{e_2}/T_{e_1} \approx 0.65$), consistent with the physical picture of semidetached secondaries as Roche-lobe-filling stars that have not yet reached thermal equilibrium with the primary. Contact systems cluster at high $T_{e_2}/T_{e_1}$ as a consequence of common-envelope energy transport. In the ($\Omega_1$, $\Omega_2$) plane, semidetached systems form a narrow locus along $\Omega_2 = \Omega_{L_1}(q)$, imposed by the Roche lobe filling condition, whereas detached and contact systems span broader ranges. 

\begin{table}[ht!]
	\centering
	\caption{Parameter distributions for the 995 training systems.}
	\label{tab:parameters}
	\begin{tabular}{lccccc}
		\hline
		\hline
		Parameter & Min & Max & Mean & Std & Median\\
		\hline
		$i$ (deg) & 38.05 & 90.00 & 82.00 & 7.56 & 83.9 \\
		$q$ ($M_2/M_1$) & 0.07 & 1.05 & 0.61 & 0.28 & 0.60\\
		$T_{e_1}$ (K) & 3160 & 60000 & 10347 & 7515 & 7175\\
		$T_{e_2}$ (K) & 3000 & 40300 & 8735 & 6793 & 6000\\
		$\Omega_1$ & 1.81 & 31.07 & 5.58 & 4.02 & 4.59\\
		$\Omega_2$ & 1.81 & 43.83 & 5.28 & 4.02 & 3.94\\
		$T_{e_2}/T_{e_1}$ & 0.11 & 1.15 & 0.86 & 0.18 & 0.95\\
		\hline
	\end{tabular}
\end{table}

Class separability was quantified through two-sample Kolmogorov--Smirnov (KS) tests for each of the five predicted parameters ($i$, $q$, $T_{e_2}/T_{e_1}$, $\Omega_1$, $\Omega_2$) between all three pairs of morphology classes (detached vs.\ semidetached, detached vs.\ contact, semidetached vs.\ contact), giving 15 tests in total. The null hypothesis in each test is that the two classes are drawn from the same univariate parameter distribution. Fourteen of the 15 tests reject the null hypothesis at $p < 0.05$ (Table~\ref{tab:ks_tests}). The single exception is the mass ratio for semidetached vs.\ contact (KS $= 0.087$, $p = 0.225$), reflecting the fact that both interacting classes are preferentially found at lower mass ratios, consistent with their shared formation channel through Roche-lobe overflow. The remaining 14 pairs show KS statistics ranging from 0.22 to 0.88 with $p$-values below $10^{-5}$, confirming that the training set provides well-separated morphological subsamples for the classifier.

\begin{table}
\centering
\caption{Pairwise two-sample KS test results for the five predicted parameters across the three morphology classes; detached (D), semidetached (SD) and contact (C). The null hypothesis ($H_0$) is the two classes are drawn from the same parameter distribution. Entries indicate rejection of $H_0$ at $p < 0.05$; the single non-significant result (SD vs C for $q$) is highlighted in italics.}
\label{tab:ks_tests}
\begin{tabular}{llcc}
\hline
Parameter & Pair & KS statistic & $p$-value \\
\hline
$i$           & D vs SD  & 0.328 & $6.7\times10^{-15}$ \\
              & D vs C   & 0.413 & $2.6\times10^{-29}$ \\
              & SD vs C  & 0.222 & $1.8\times10^{-6}$ \\
$q$           & D vs SD  & 0.743 & $1.2\times10^{-81}$ \\
              & D vs C   & 0.741 & $4.0\times10^{-101}$ \\
              & \textit{SD vs C}  & \textit{0.087} & \textit{0.225} \\
$T_{e_2}/T_{e_1}$    & D vs SD  & 0.684 & $6.2\times10^{-68}$ \\
              & D vs C   & 0.219 & $2.0\times10^{-8}$ \\
              & SD vs C  & 0.775 & $5.7\times10^{-83}$ \\
$\Omega_1$    & D vs SD  & 0.571 & $5.4\times10^{-46}$ \\
              & D vs C   & 0.636 & $4.5\times10^{-72}$ \\
              & SD vs C  & 0.388 & $4.1\times10^{-19}$ \\
$\Omega_2$    & D vs SD  & 0.879 & $3.8\times10^{-123}$ \\
              & D vs C   & 0.645 & $2.9\times10^{-74}$ \\
              & SD vs C  & 0.267 & $3.3\times10^{-9}$ \\
\hline
\end{tabular}
\end{table}

\begin{figure}[ht!]                                                
\centering
\epsscale{0.6}                                                     
\plotone{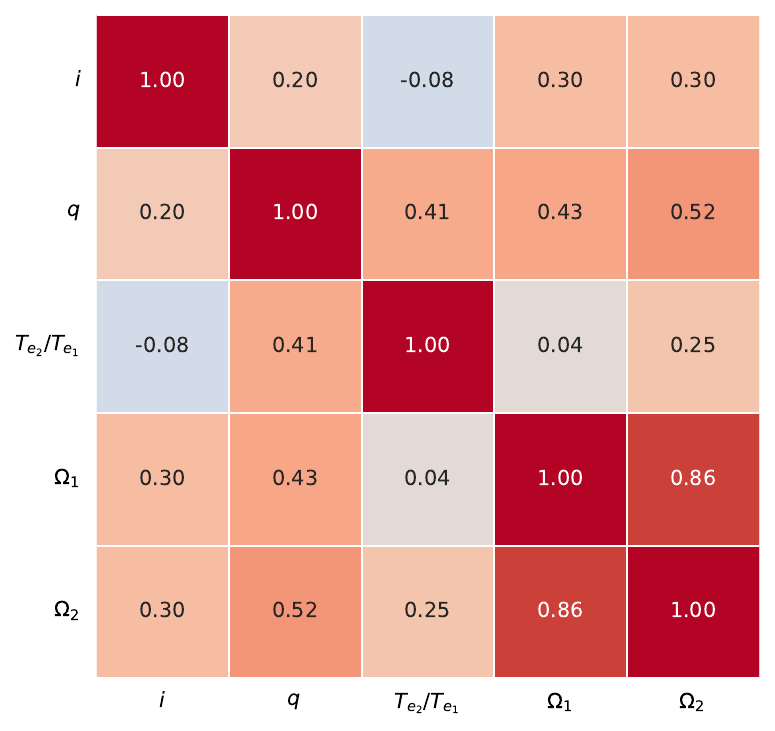}
\caption{Pearson correlation matrix of the five predicted parameters for the 995 training systems.}
\label{fig:param_correlation} 
\end{figure}

We calculated the Pearson correlation coefficients between the physical attributes (Fig.~\ref{fig:param_correlation}) to evaluate the independence of the parameter space. While the inclination remains largely independent of other parameters, we observe significant collinearity between the potentials. As discussed by \citet{prs05}, this structure reflects the astrophysical priors used to generate the dataset and underscores the parameter degeneracies characteristic of the Roche geometry that the model must resolve. While $\Omega_1$ and $\Omega_2$ are highly correlated, ensemble tree methods like XGB are robust to multicollinearity. Unlike linear regression, decision trees naturally handle redundant features by selecting the single most discriminative feature at each split, preventing parameter instability.

Given these complexities, we strictly verified the integrity of our training pipeline and confirmed the absence of data leakage. Feature standardization using \texttt{StandardScaler} \citep{ped11} was performed exclusively on training folds and subsequently applied to validation folds. Although ensemble tree methods such as RF and XGB are invariant to monotonic feature transformations and do not require scaling for prediction, \texttt{StandardScaler} was applied for two reasons. First, the 51 features span several orders of magnitude in dynamic range, and the PCA dimensionality reduction experiment (Sec.~\ref{sec:reg_per}) requires standardised inputs to produce unbiased principal components. Therefore, without scaling, high-variance features would dominate the PCA projections irrespective of their physical relevance. Second, a unified preprocessing pipeline for both the PCA analysis and the ensemble model training ensures internal consistency across all experiments. Cross-validation splits ensured strictly non-overlapping training and validation indices across all five folds.
\end{sloppypar}

\subsection{Feature Engineering}
\label{sec:features}

We extract 51 domain-specific features from each phase-folded light curve, organized into four categories designed to capture physically meaningful information about eclipsing binary morphology and component properties. Eight features characterize the global light curve properties: baseline flux (median), minimum, maximum, mean, standard deviation, range, skewness, and kurtosis. These statistics provide information about overall variability amplitude and flux distribution asymmetry. Table~\ref{tab:features} summarizes the feature categories and counts. Once the global statistical properties were established, we focused on extracting specific morphological markers. The most critical of these is the eclipse geometry; therefore, our eclipse detection employs a dynamic algorithm that identifies minima by relative depth rather than assuming fixed phase positions. The deepest minimum defines the primary eclipse; a second search with a $\pm0.15$ phase mask around the primary identifies the secondary eclipse. Four depth and width features are computed: primary and secondary eclipse depths (measured from the median baseline) and their corresponding widths (fraction of the light curve below half-depth). Three additional features quantify the O'Connell effect, the brightness difference between maxima at quadrature phases, which serves as an indicator of effective temperature asymmetries common in contact binaries. These features include the absolute brightness difference between maxima, the range, and standard deviation. 

\begin{table}[ht!]
\centering
\small
\caption{Summary of extracted features by category. The complete feature list is provided at \protect\url{https://github.com/burakulas/ml-ebs/blob/main/Feature_List.md}}
\label{tab:features}
\begin{tabular}{l c l}
\hline
\hline
Category & Features & Description \\
\hline
Basic statistics & 8 & Global light curve properties \\
Morphological features & 7 & Eclipse depths and widths, asymmetry in maximum\\
Fourier components & 17 & Harmonic amplitudes, phases, ratios \\
Phase-binned statistics & 19 & Binned means and consecutive differences \\
\hline
Total & 51 & \\
\hline
\end{tabular}
\end{table}

Fourier decomposition captures the periodic structure of the light curve through 17 features. We compute amplitudes for the first 10 harmonics, phases for harmonics 1-3, amplitude ratios of harmonics 2, 3, and 4 relative to the fundamental, and total Fourier power. The harmonic amplitudes describe the signal power distribution; the second harmonic is particularly diagnostic, as contact binaries with ellipsoidal variations show strong second-harmonic contributions \citep{ruc93}, while detached systems with narrow eclipses exhibit power distributed across higher harmonics \citep{poj02}. Complementarily, the phases of the lowest three harmonics are included to encode light curve asymmetries and eclipse timing offsets commonly associated with the O'Connell effect or orbital eccentricity.

The phase interval $[0.25,1.25]$ is divided into 10 equal bins of width 0.1 phase units, with centers ranging from 0.30 to 1.20. For each bin, we compute the mean flux, yielding 10 features. Additionally, 9 consecutive difference features ({\it{bin k+1}} minus {\it{bin k}}) capture eclipse ingress and egress slopes, which encode information about stellar radii and orbital inclination. The feature set is designed with physical interpretability in mind. Eclipse depth ratios directly encode the effective temperature ratio $(T_{e_2}/T_{e_1})^4$ through the Stefan-Boltzmann law, independent of inclination for total eclipses. Eclipse widths depend on the relative stellar radii and orbital geometry. The O'Connell effect magnitude correlates with degree of contact and effective temperature inhomogeneities. Fourier amplitude distributions distinguish morphological classes: contact systems show smooth, sinusoidal variations dominated by low harmonics, while detached systems exhibit sharp eclipses requiring higher-order terms. 

\section{Model Architecture and Training}
\label{sec:training}

\subsection{Training Methodology}\label{sec:train_method}
\begin{sloppypar}
We employ a multi-task machine learning framework\footnote{\label{fn:github}\url{https://github.com/burakulas/ml-ebs}} that performs both morphology classification and physical parameter estimation from the 51 engineered features. RF and XGB ensemble machine learning algorithms were trained for comparison. Both methods are chosen for their ability to handle non-linear relationships, provide feature importance rankings, and demonstrate robust performance on tabular data with mixed feature types. RF serves as our interpretable baseline. As a bagging ensemble method, it constructs multiple decision trees trained on bootstrap samples and averages their predictions to reduce variance. The majority voting mechanism for classification and mean prediction for regression make RF predictions straightforward to interpret. XGB represents a state-of-the-art gradient boosting framework. Unlike RF's independent tree construction, XGB builds trees sequentially, with each tree correcting errors from previous iterations. Regularization terms prevent overfitting, while the gradient-based optimization enables faster convergence and often superior predictive performance. 

\begin{figure*}
	\centering
	\plotone{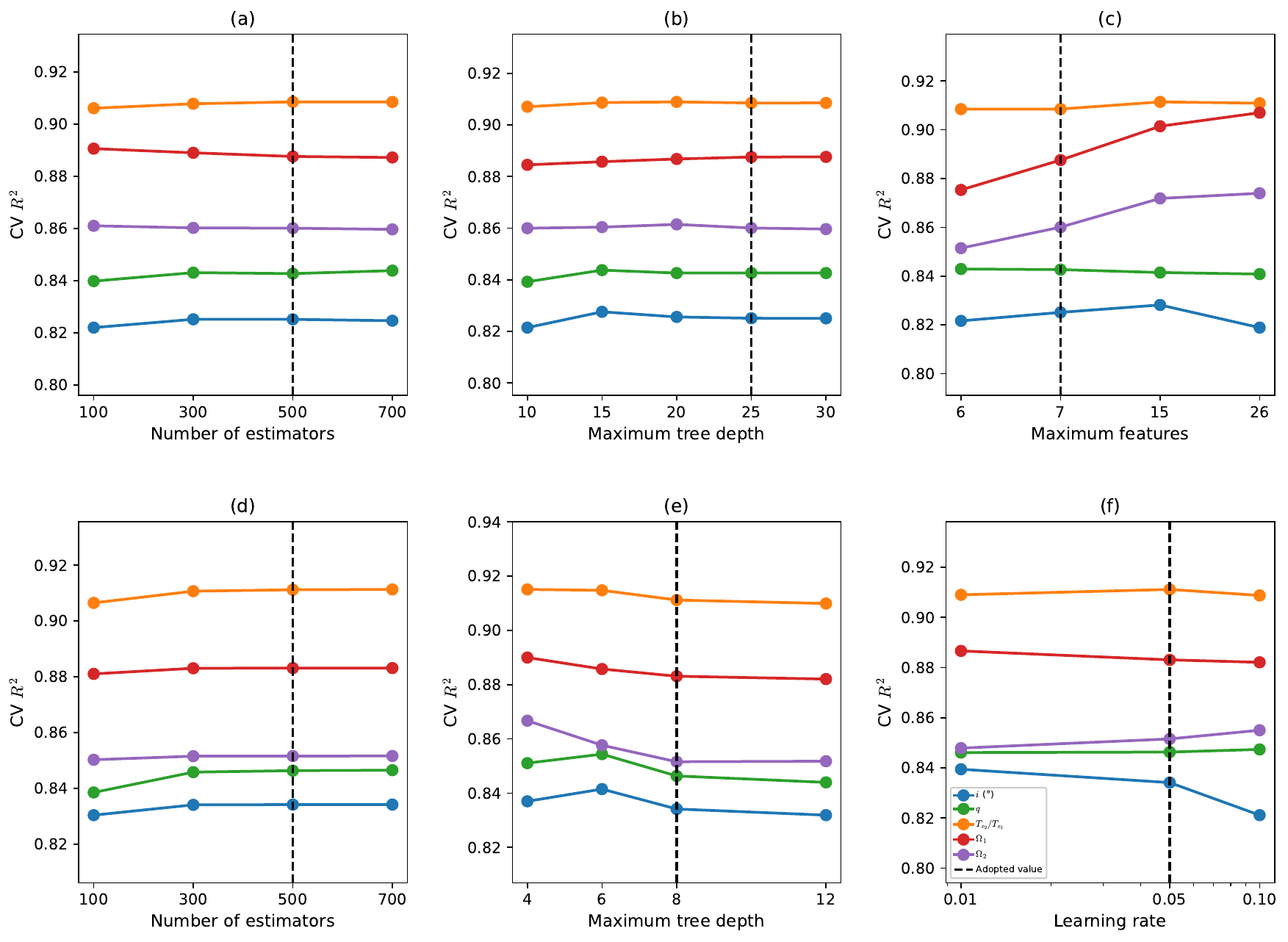}
	\caption{Top panel:  RF hyperparameter sensitivity analysis on the cross-validation set. The plot show CV $R^2$ variation with the number of estimators (a), the maximum tree depth (b), and the maximum features per split (c). For panel (c), the numbers on the x-axis correspond to $\log_{2} (51) \approx 6$, $\sqrt{51}\approx 7$, $0.3\times 51\approx 15$, and $0.5\times 51\approx 26$ features considered at each split out of the 51 total features. Bottom panel: XGB hyperparameter sensitivity analysis on the same set. In each panel, one hyperparameter is varied while the other two are held at their adopted values (dashed vertical line); the five curves show the per-parameter CV $R^2$ for $i$, $q$, $T_{e_2}/T_{e_1}$, $\Omega_1$, and $\Omega_2$. (d) Number of estimators, (e) maximum tree depth, (f) learning rate.}
	\label{fig:hyperparams}
\end{figure*}

To determine the hyperparameter choices, we performed a systematic sensitivity analysis on the cross-validation set (Fig.~\ref{fig:hyperparams}). For the RF model, the number of estimators has negligible effect on performance, with the mean $R^2$ varying by only 0.001 across the tested range; 500 trees are sufficient. For maximum tree depth (Fig.~\ref{fig:hyperparams}b), $R^2$ varies by less than 0.003 between depths 10 and 30, with performance stabilising above depth 15 which is consistent with RF's inherent resistance to overfitting through bootstrap aggregation. The most influential RF hyperparameter is the number of features considered at each split (Fig.~\ref{fig:hyperparams}c): using 30\% of features ($\approx$15) yields a mean $R^2$ of 0.871, marginally higher than the adopted setting ($\approx$7 features, $R^2 = 0.865$). We retain maximum features equals to square root of the number of features ($\sqrt{51}$) as it is the established default for RF regression that maximises tree decorrelation and ensemble diversity; the 0.6\% difference falls within cross-validation variability. We also varied each of the three primary XGB hyperparameters across a focused grid while holding the remaining parameters at their adopted values. For the number of estimators (Fig.~\ref{fig:hyperparams}d), the mean cross-validation $R^2$ varies by only 0.004; performance plateaus at 500 with no measurable gain at 700, confirming 500 as the point of diminishing returns. $R^2$ of maximum tree depth (Fig.~\ref{fig:hyperparams}e),  varies by 0.008, with shallow trees such as 4 slightly outperforming deeper ones, indicating that the chosen depth of 8 lies within a stable plateau. For learning rate (Fig.~\ref{fig:hyperparams}f), $R^2$ varies by only 0.003; the adopted value of 0.05 provides stable convergence without the over-compression seen at 0.01 or the instability at 0.1. Across the complete 48-combination grid of maximum tree depth and learning rate, the mean $R^2$ ranges from 0.859 to 0.874 (a total spread of 1.7\%), confirming robustness of hyperparameter choice. 
 
Based on this analysis, we configure RF models with 500 trees (\texttt{n\_estimators=500}) and maximum depth of 25 (\texttt{max\_depth=25}), allowing the ensemble to capture complex non-linear relationships between features and stellar parameters. We set \texttt{min\_samples\_split=2} and \texttt{min\_samples\_leaf=1} to permit fine-grained decision boundaries, while \texttt{max\_features='sqrt'} ($\sqrt{51} \approx 7$ features per split) introduces randomness that improves ensemble diversity and reduces overfitting. A fixed random seed (\texttt{random\_state=42}) ensures reproducibility. XGB regression models employ 500 trees with moderate depth (\texttt{max\_depth=8}) and low learning rate (\texttt{learning\_rate=0.05}), implementing a conservative boosting strategy that prevents overfitting. We subsample 80\% of both training instances (\texttt{subsample=0.8}) and features (\texttt{colsample\_bytree=0.8}), adding stochasticity similar to RF's bootstrap approach. L1 and L2 regularization terms (\texttt{reg\_alpha=0.1}, \texttt{reg\_lambda=0.1}) penalize model complexity, favoring simpler decision rules that generalize better to unseen data.For morphology classification, we adjust hyperparameters to the discrete 3-class problem. The RF classifier uses 500 trees with \texttt{max\_depth=20}, \texttt{min\_samples\_split=5}, and \texttt{min\_samples\_leaf=2}. The XGB classifier employs 500 trees with \texttt{max\_depth=6}, \texttt{learning\_rate=0.05}, \texttt{objective='multi:softmax'}, and \texttt{num\_class=3}. The reduced tree depths and increased minimum sample requirements prevent overfitting on the classification task, which has lower complexity than continuous parameter regression.

\end{sloppypar}

We employed a 5-fold stratified cross-validation scheme to ensure that our performance estimates are robust against morphological imbalances, while maintaining morphology class proportions in each fold. The 845 training systems are divided into 5 folds ($\sim$169 systems each), stratified by morphology class (detached, semidetached, contact). For each fold, we use 4 folds as the training set ($\sim$676 systems, 80\%) and the remaining fold as the validation set ($\sim$169 systems, 20\%). Models are trained on the training set and evaluated on the held-out validation set, with the process repeating until each fold has served as the validation set once. The stratification ensures balanced morphology representation across all folds. Each fold contains approximately 42.3\% detached, 34.2\% contact, and 23.5\% semidetached systems, matching the overall dataset distribution. A fixed random seed ensures reproducibility across all experiments. Critically, the \texttt{StandardScaler} is fit independently for each fold using only the training partition, preventing data leakage from validation sets into model training. Internal verification confirmed strict separation of training and validation indices to prevent data leakage.

\begin{figure*}[ht!]
\centering 
\plotone{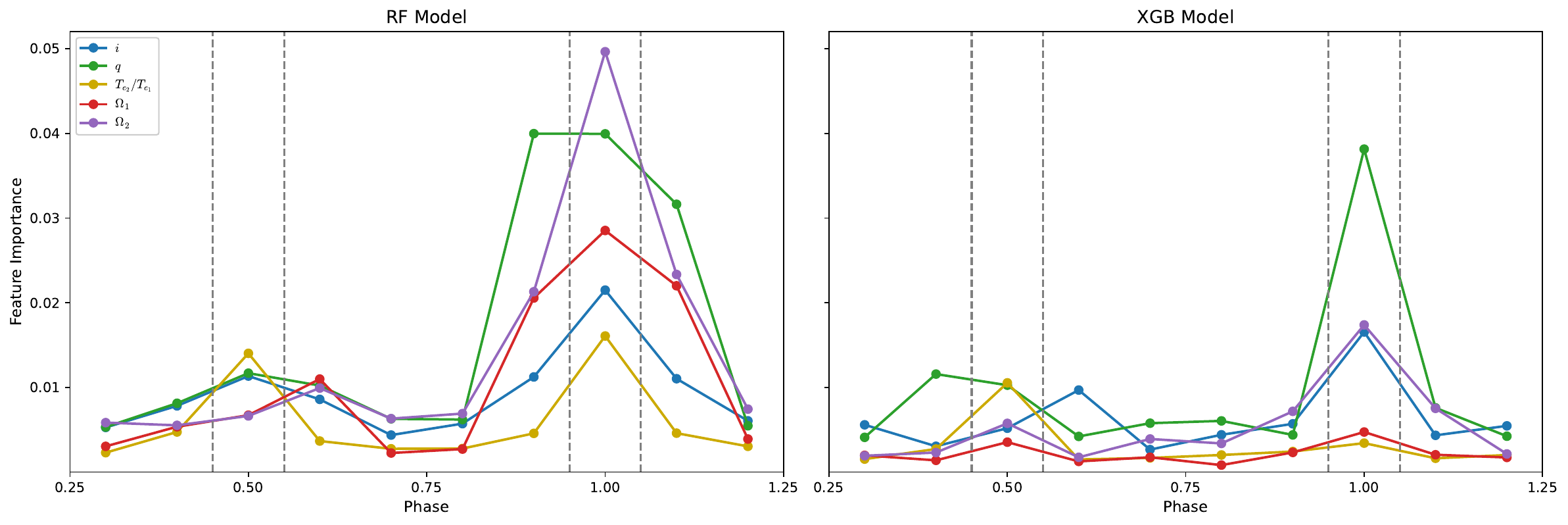}
\caption{Importance of phase-binned statistics for RF (left) and XGB (right) models. Regions covered by vertical dashed lines mark secondary and primary eclipse phases.}
\label{fig:phase_importance}                                     
\end{figure*}

Our framework trains separate models for each task: five independent regression models predicting $i$, $q$, $T_{e_2}/T_{e_1}$, $\Omega_1$, and $\Omega_2$, plus one classification model predicting morphology (detached, semidetached or contact). All models share the same 51-dimensional feature representation but learn task-specific patterns. This architecture offers several advantages over multi-output models. First, task independence ensures that each parameter's prediction errors do not propagate to other parameters. Second, parameters with different prediction difficulties (e.g., $T_{e_2}/T_{e_1}$ vs. $i$) can utilize different effective model capacities. Third, feature importance can be analyzed separately for each parameter, revealing which light curve characteristics inform specific stellar properties. This analysis confirms that the models effectively identify the physical drivers of light curve morphology. As shown Fig.~\ref{fig:phase_importance}, the feature importance distribution for all parameters peaks distinctly at the primary and secondary minima, indicating that the ensemble trees correctly prioritize these regions as the most information-rich. The predictive focus varies by parameter in an astrophysically consistent manner. For instance, the effective temperature ratio shows significant importance at the secondary minimum phase for XGB model, where the relative flux depth directly constrains the surface brightness ratio. Fourth, models can be trained, evaluated, and improved independently without retraining the entire system. For each parameter, both RF and XGB models are trained across all 5 folds, yielding 5 trained models per algorithm per parameter (50 regression models total) plus 5 classification models per algorithm (10 classification models total). The complete training code and corresponding data are available online\footref{fn:github}.

\begin{figure}
\centering
\epsscale{0.7}
\plotone{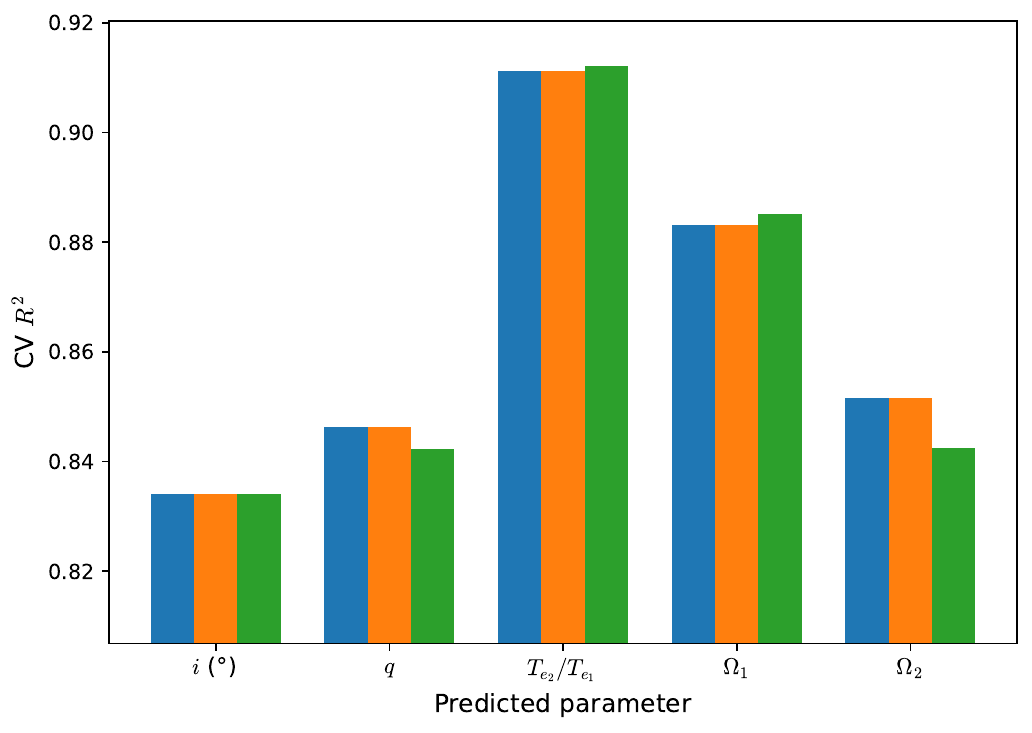}
\caption{Comparison of three XGB-based regression architectures evaluated on the set of 5-fold stratified CV, 845 systems). Blue bars correspond to five independent single-output \texttt{XGBRegressor} models (mean $R^2 = 0.867$), a \texttt{MultiOutputRegressor} wrapper that fits one independent \texttt{XGBRegressor} per target represented by orange bars ( mean $R^2 = 0.867$). \texttt{RegressorChain} of \texttt{XGBRegressor} models (green bars, mean $R^2 = 0.863$), in which predictions of earlier outputs are appended as features for later ones with the chain order of $i$, $T_{e_2}/T_{e_1}$, $q$, $\Omega_1$, $\Omega_2$).}
\label{fig:multioutput}
\end{figure}

To directly validate these architectural advantages, we compared the single-output approach against two multi-output strategies on the cross-validation set (Fig.~\ref{fig:multioutput}). A \texttt{MultiOutputRegressor} wrapper, which trains one independent estimator per output and is therefore structurally equivalent to our approach, and a \texttt{RegressorChain}, which appends predictions from earlier outputs as features for subsequent outputs along the chain $i \rightarrow T_{e_2}/T_{e_1} \rightarrow q \rightarrow \Omega_1 \rightarrow \Omega_2$. All three methods use identical XGB hyperparameters and are evaluated on the same 5-fold stratified cross-validation splits. The mean cross-validation $R^2$ for the single-output \texttt{XGBRegressor}, \texttt{MultiOutputRegressor}, and \texttt{RegressorChain} architectures is 0.867, 0.867, and 0.863, respectively. The \texttt{MultiOutputRegressor} result (identical to single-output to four decimal places) confirms the structural equivalence of the two approaches. The \texttt{RegressorChain} performs slightly worse overall ($\Delta \bar{R}^2 = -0.002$), with the largest degradation occurring for $\Omega_2$ ($\Delta R^2 = -0.009$), the final parameter in the chain. This parameter receives the accumulated prediction errors of all four preceding outputs as additional input features, directly demonstrating the error-propagation mechanism that the single-output architecture avoids. We note that this comparison was conducted with XGB; for RF, the single-output and \texttt{MultiOutputRegressor} architectures are       mathematically identical, as bagging trains each tree independently without sequential dependence between outputs. The \texttt{RegressorChain} degradation mechanism is specific to sequential learners and does not apply to RF's parallel bootstrap aggregation.

\begin{figure}[ht!]
\centering
\plotone{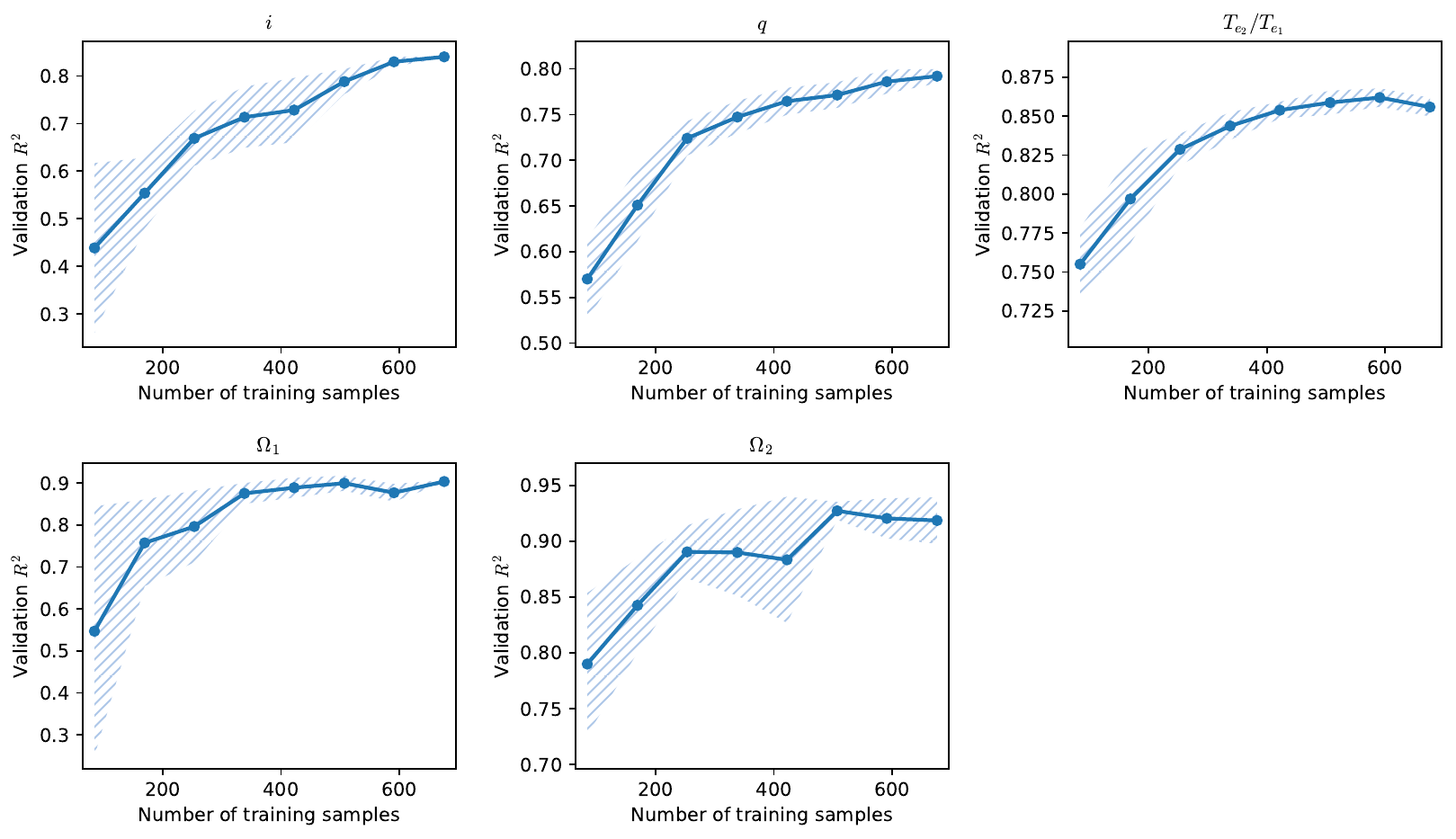}
\caption{Learning curves for XGB regression across five stellar parameters.	Validation $R^2$ is shown as a function of training set size (10\%--80\% of the 845-system cross-validation pool. Each point represents the mean over five independent random subsamples; shaded bands indicate $\pm$1 standard deviation.	The held-out set was excluded throughout.}
\label{fig:lr_curves}
\end{figure}

\begin{figure}[ht!]
\centering
\plotone{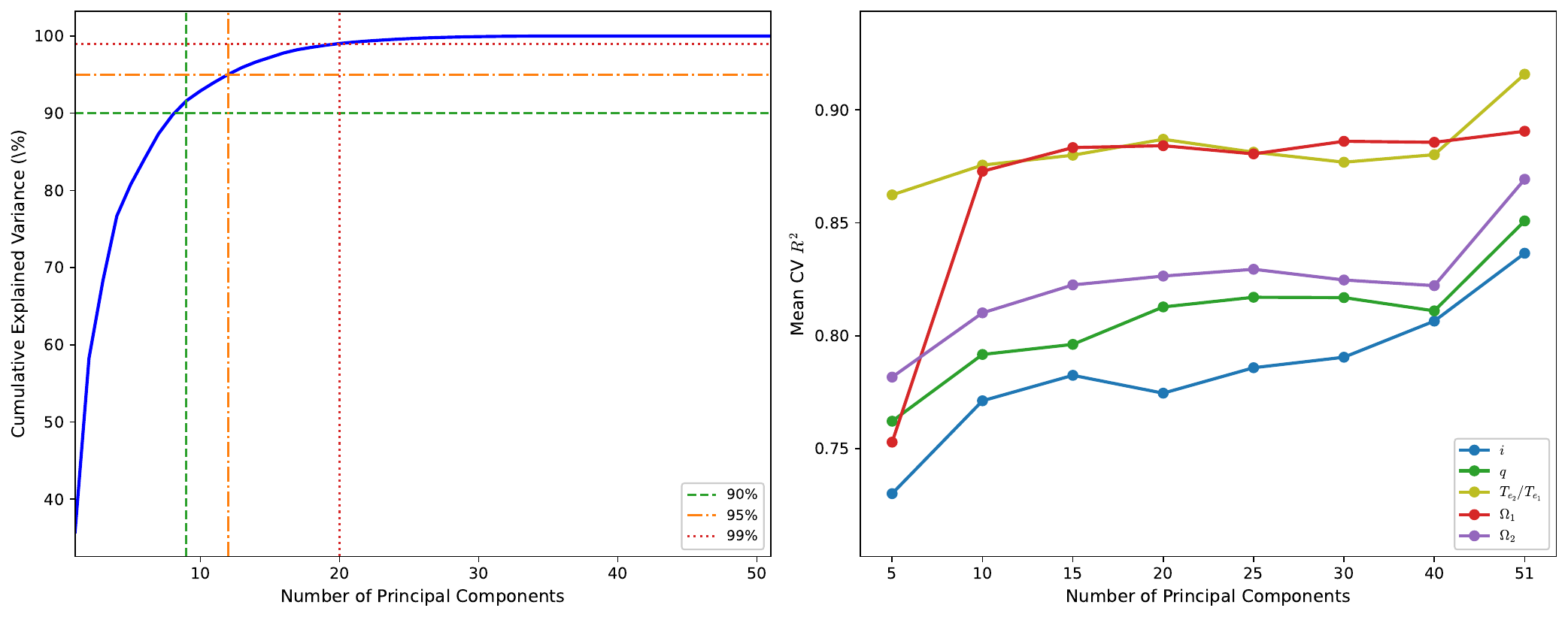}
\caption{PCA dimensionality reduction experiment. Left: cumulativeexplained variance as a function of the number of principal components, with dashed lines marking the 90\% (9 components), 95\% (12 components), and 99\% (20 components) thresholds. Right: mean 5-fold CV $R^2$ on the 845-system CV set as a function of the number of principal components for each parameter.} 
\label{fig:pca_exp}
\end{figure}

\subsection{Training Results}
\label{sec:results}
\subsubsection{Regression Performance}\label{sec:reg_per}

The RF and XGB models achieve comparable cross-validation performance (mean $R^2$ of 0.865 and 0.867, respectively), with XGB showing marginal advantages for $i$, $q$, and $T_{e_2}/T_{e_1}$, and RF performing equivalently for the surface potentials. We adopt XGB as the primary prediction model for two reasons. First, XGB natively supports GPU acceleration, enabling the rapid inference required for large-scale survey application (Sec.~\ref{sec:predictions}), whereas RF relies exclusively on CPU parallelism. Second, XGB achieves lower MAE (Mean Absolute Error) on the majority of parameters. The RF ensemble is retained for its unique capability of providing per-sample uncertainty estimates through the tree-prediction distribution. The cross-validated and held-out test performance of XGB models on the five stellar parameters is presented in Table~\ref{tab:xgboost_regression}. Although training $R^2$ approaches unity across all training sizes, which is an expected behavior for deep gradient-boosted ensembles with  sufficient tree depth, several lines of evidence confirm that this does not reflect overfitting. The learning curves (Fig.~\ref{fig:lr_curves}) show  validation $R^2$ rising steadily and converging toward training $R^2$, with no divergence at larger training sizes. The mean absolute difference between cross-validation and held-out $R^2$ across the five parameters is only 0.052, and for three of five parameters the held-out $R^2$ actually exceeds the cross-validation value (Table~\ref{tab:xgboost_regression}). The held-out set (150 systems) was separated prior to any training or hyperparameterselection, providing a fully independent evaluation. These results, combined with the L1/L2 regularization and stochastic subsampling described in Sec.~\ref{sec:train_method}, confirm that the models generalize beyond the training data. The learning curves (Fig.~\ref{fig:lr_curves}) show validation $R^2$ increasing steadily with training size and approaching a plateau by 70–80\% of the cross-validation set for most parameters. The standard deviation of $R^2$ across five random subsamples narrows substantially, indicating stable convergence at the full dataset size. Mass ratio shows the slowest convergence, consistent with its inherent geometric degeneracy in photometric data. The PCA experiment (Fig.~\ref{fig:pca_exp}) demonstrates that prediction accuracy increases monotonically from 5 to 51 principal components for every parameter, reaching its optimum only at the full feature representation. The 9 components capturing 90\% of feature variance yield substantially lower performance than the full set, confirming that each feature category (eclipse morphology, Fourier harmonics, and phase-binned statistics) encodes complementary physical information.

The held-out test yields $R^2$ values broadly consistent with cross-validation (Table~\ref{tab:xgboost_regression}), providing direct confirmation that the model generalizes without overfitting. The effective temperature ratio $T_{e_2}/T_{e_1}$ achieves the highest cross-validation $R^2$ (0.911), indicating that the eclipse-specific features effectively capture the depth ratio that directly relates to     effective temperature differences. The surface potential parameters $\Omega_1$ and $\Omega_2$ demonstrate strong and stable performance, with held-out  $R^2$ exceeding their cross-validation values. The mass ratio achieves cross-validation $R^2$ = 0.85 but shows the largest held-out degradation ($R^2$ = 0.77), as expected from the well-known $q$--$i$ ambiguity discussed above. Inclination yields cross-validation $R^2$ = 0.83 but improves on the held-out set ($R^2$= 0.89), suggesting that the cross-validation folds underestimate performance for this well-constrained geometric parameter.

\begin{table}[ht!]
\centering
\small
\caption{XGB regression performance. CV columns report mean ± standard deviation across 5-fold cross-validation on the 845-system training pool. Held-out columns report metrics on the test set (see Sec.~\ref{sec:results}). MAE units are degrees for $i$, dimensionless for $q$, $T_{e_2}/T_{e_1}$ and $\Omega_1$ and $\Omega_2$. RMSE is root mean square error.}
\label{tab:xgboost_regression}
\begin{tabular}{lccccc}
\hline
Parameter & CV $R^2$ (mean $\pm$ std) & CV MAE & Held-out $R^2$ & Held-out MAE & Held-out RMSE \\
\hline
$i (deg)$ & 0.834 $\pm$ 0.048 & 1.90 $\pm$ 0.13 & 0.887 & 1.534 & 2.269 \\
$q$ 	   & 0.852 $\pm$ 0.021 & 0.070 $\pm$ 0.002 & 0.765 & 0.075 & 0.130 \\
$T_{e_2}/T_{e_1}$ & 0.911 $\pm$ 0.010 & 0.032 $\pm$ 0.001 & 0.882 & 0.036 & 0.067 \\
$\Omega_1$ & 0.880 $\pm$ 0.074 & 0.58 $\pm$ 0.08 & 0.911 & 0.455 & 0.928 \\
$\Omega_2$ & 0.859 $\pm$ 0.078 & 0.59 $\pm$ 0.13 & 0.920 & 0.466 & 0.902 \\
\hline
\end{tabular}
\end{table}

\begin{figure*}[ht!]
\centering
\plotone{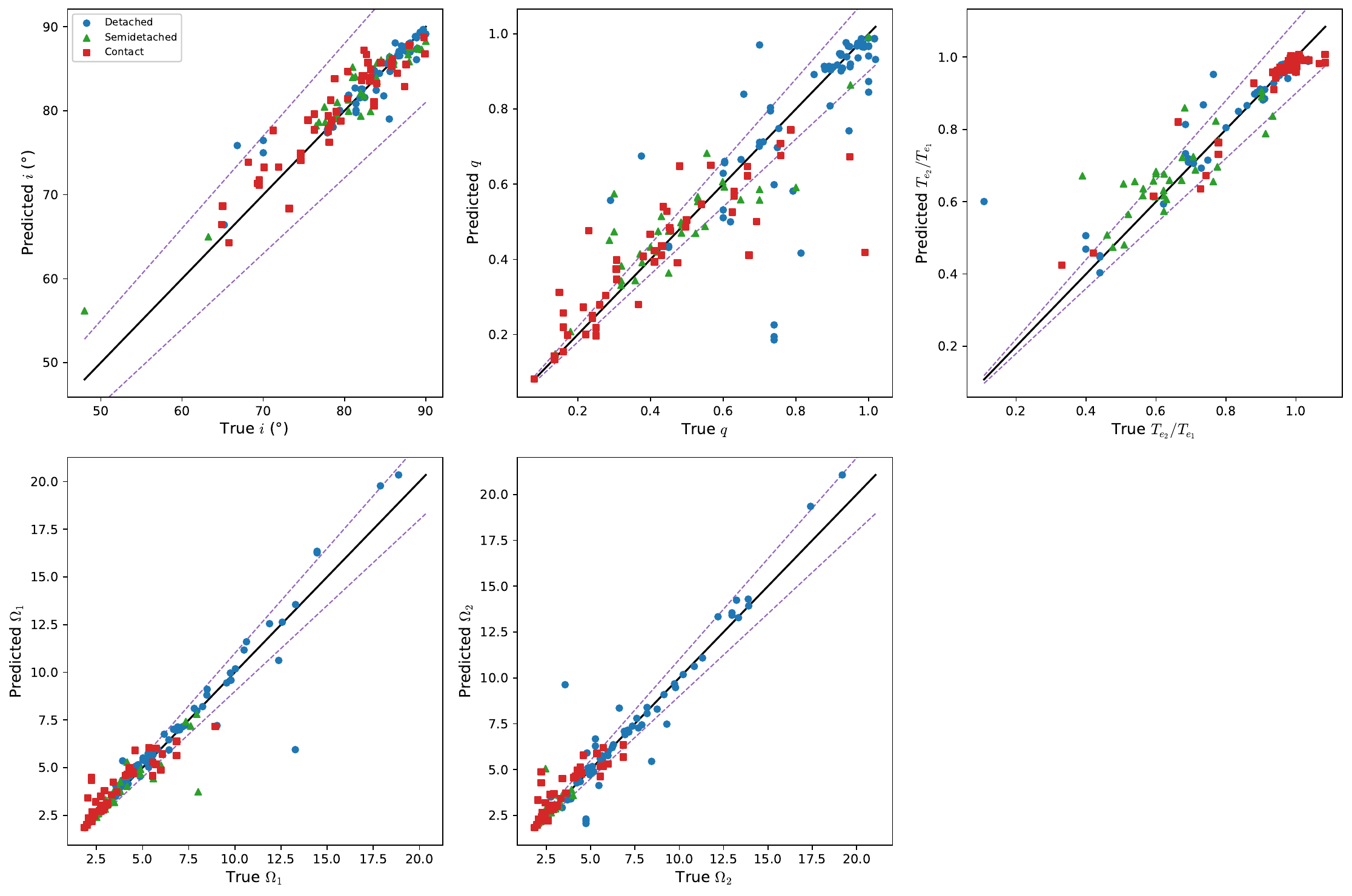}
\caption{Predicted versus true values for XGB regression on the held-out test set. Diagonal black lines indicate perfect agreement while the purple dashed lines mark $\pm$10\% relative error bounds.}
\label{fig:predicted_vs_actual}
\end{figure*}

\begin{figure*}[ht!]
\centering
\plotone{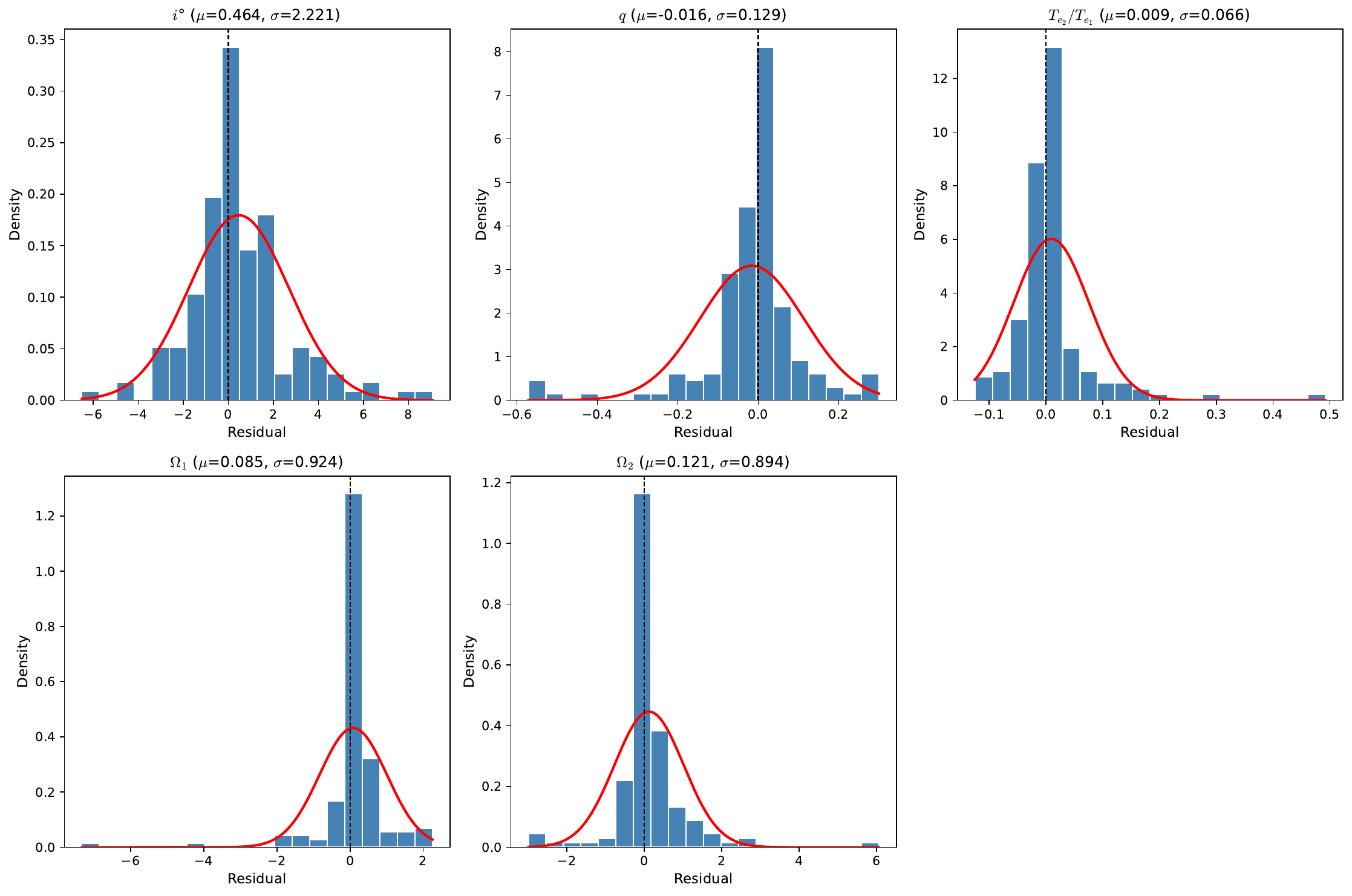}
\caption{XGB residual distributions (predicted minus true) for the held-out test set. Histograms show  residual density per parameter; red curves show fitted Gaussian distributions. Panel headers report mean ($\mu$) and standard deviation ($\sigma$) of each distribution.}
\label{fig:residuals}
\end{figure*}

The model predictions track the ground truth tightly across the entire parameter space, as evidenced by the strong diagonal alignment in  Fig.~\ref{fig:predicted_vs_actual}. Notably, the effective temperature ratio panel shows tight clustering along the ideal prediction line,  consistent with its high $R^2$ value. The surface potentials $\Omega_1$ and $\Omega_2$ demonstrate similarly strong agreement, with the     highest held-out $R^2$ values (0.911 and 0.920, respectively). The mass ratio panel exhibits broader scatter, particularly at low $q$,  reflecting the limited photometric sensitivity to $q$ at low mass ratios. The inclination panel reveals increased scatter at lower values ($i<70~\deg$), reflecting the training set's natural bias toward high orbital inclinations: edge-on systems produce deeper, more detectable eclipses as discussed in Sec.~\ref{sec:dataset}. While the current framework performs robustly for the majority of detected eclipsing binaries, the predictive uncertainty increases for grazing systems. Future iterations of this work will address this regime by incorporating independent geometric parameters derived from spectroscopic and radial velocity observations, thereby providing tighter constraints for non-edge-on geometries. Fig.~\ref{fig:residuals} presents residual distributions for each parameter, demonstrating approximately Gaussian error distributions centered near zero, which validates the regression assumptions. A small systematic overestimation is present for inclination ($\mu=+0.46^o$, corresponding to $~0.2\sigma$), likely reflecting the training set's concentration at high inclinations, while the remaining parameters show negligible bias ($\lvert\mu \rvert < 0.13$).

\begin{figure*}
\centering
\includegraphics[scale=0.5]{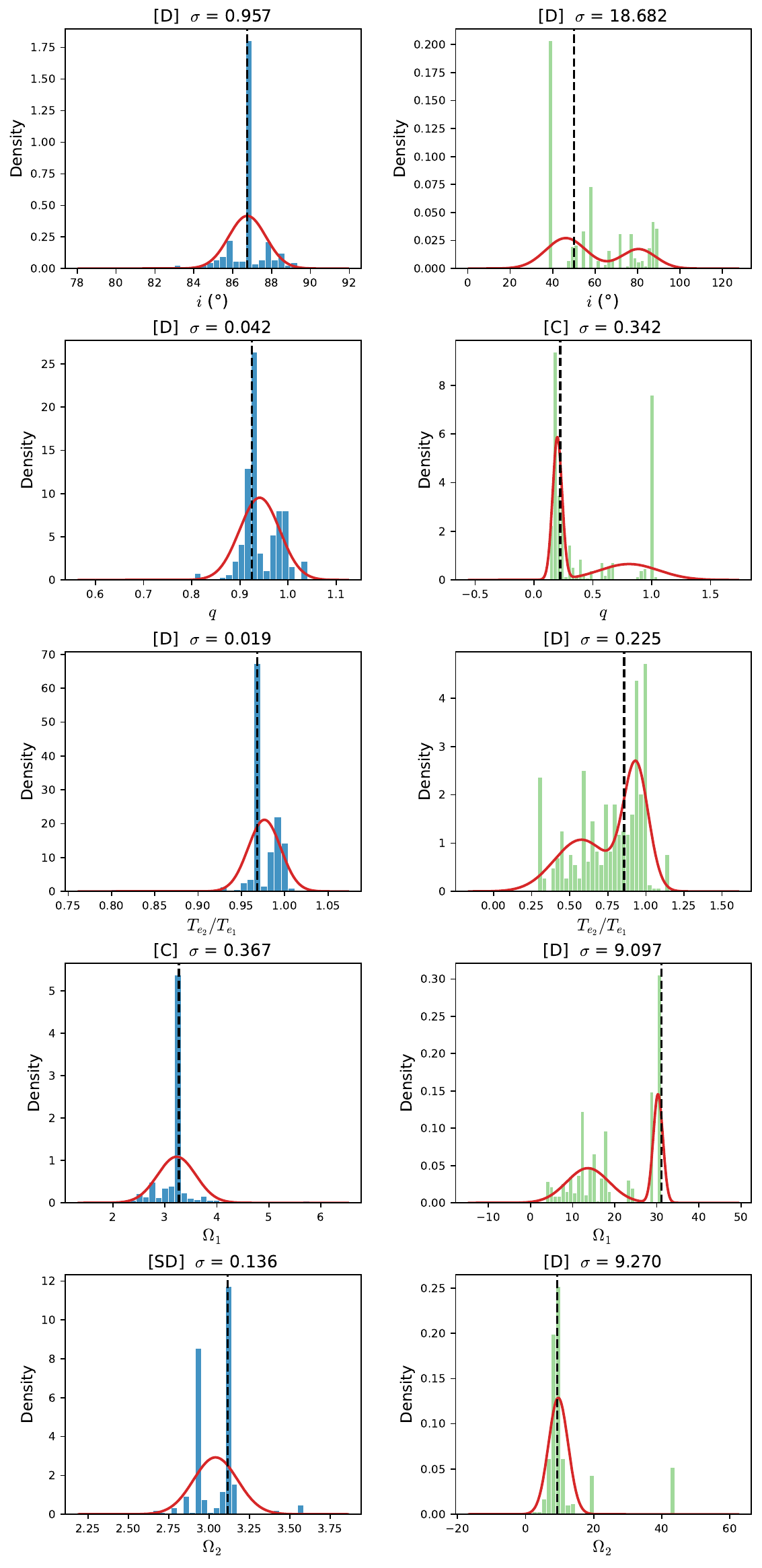}
\caption{Representative RF tree-prediction distributions for all five predicted parameters. For each cross-validation sample, predictions from all 500 individual decision trees are collected out-of-fold and a Gaussian Mixture Model (GMM with 1 or 2 components, selected by BIC) is fitted to the resulting histogram. Each panel  shows a single representative system; the morphology label in brackets (D,SD,C) indicates the class of that particular star: detached, semidetached or contact. These individual examples are selected to illustrate the range of predictive certainty rather than to represent any particular morphology class. The left column shows the tightest unimodal case (1-component GMM) for each parameter, representing a well-constrained prediction. The right column demonstrates the sample with the largest GMM standard deviation (2-component GMM), representing the highest prediction uncertainty. The dashed black line marks the true parameter value while the solid red curve is the fitted GMM.}
\label{fig:rf_uncertainty}
\end{figure*}

To characterize the probabilistic uncertainty of individual RF predictions, we followed the approach of \citet{dis18} and extracted the full distribution of predictions from all 500 decision trees in each RF fold model. This approach is inherently specific to bagging-based ensembles such as RF, whose trees are trained independently and each produces a complete target prediction. In gradient-boosted ensembles such as XGB, individual trees are sequential residual correctors whose outputs sum to the final prediction. Therefore, the distribution of individual tree outputs has no direct probabilistic interpretation. For each cross-validation sample (out-of-fold only and no tree had access to the sample during training), we fitted Gaussian Mixture Models (GMM) to the tree-prediction histogram, selecting the number of components via the Bayesian Information Criterion (BIC). The effective standard deviation of the fitted GMM provides a per-sample uncertainty estimate; median values across the 845 cross-validation samples are $\tilde{\sigma}_i = 3.26$\degr, $\tilde{\sigma}_q = 0.133$, $\tilde{\sigma}_{T_2/T_1} = 0.057$,  $\tilde{\sigma}_{\Omega_1} = 1.17$, and $\tilde{\sigma}_{\Omega_2} = 1.11$, of comparable magnitude to the XGB cross-validation RMSE values (Table~\ref{tab:xgboost_regression}), indicating consistent error scales across both ensemble methods. Individual distributions range from tight and near-Gaussian to broad, and in some cases clearly bimodal, reflecting the varying degrees of photometric constraint across different systems. Contact systems show systematically larger spread ($\tilde{\sigma}_i = 4.94$\degr) than detached systems ($\tilde{\sigma}_i = 3.03$\degr), consistent with the smooth, continuously varying light curves of contact configurations, which lack the sharp eclipse features needed to tightly constrain inclination. Representative tight high-uncertainty examples for all five predicted parameters are shown in Fig.~\ref{fig:rf_uncertainty}. The fact that this distributional uncertainty analysis is not available for gradient-boosted ensembles further motivates the dual-model architecture. XGB serves as the primary regression model for its  GPU-accelerated inference and marginally lower MAE, while the RF ensemble provides calibrated per-sample uncertainty information.

\subsubsection{Classification Performance}

Morphology classification achieves excellent performance with XGB obtaining 95.38\% $\pm$ 1.15\% overall accuracy and macro-averaged F1 score of 94.87\% $\pm$ 1.25\% across 5-fold cross-validation. The weighted F1 score closely matches the overall accuracy, indicating consistent performance across all morphology classes. RF achieves nearly identical performance (accuracy: 95.50\%, F1 macro: 94.92\%), demonstrating that the morphological distinction between detached, semidetached, and contact binaries is robustly encoded in the 51 eclipse and Fourier features regardless of the specific ensemble algorithm employed.

Per-class classification metrics, revealing differential performance across morphologies are presented in Table~\ref{tab:classification}. Contact and detached systems demonstrate comparable high performance. Detached systems exhibit the highest precision, benefiting from their larger sample size and distinctive light curve characteristics with well-separated, sharp eclipses. Contact systems achieve the highest F1 score despite having fewer examples; their continuous light variation creates a unique signature that minimizes false positives. Semidetached systems show lower metrics, characterized specifically by lower precision relative to recall. This discrepancy suggests that while the model successfully identifies most semidetached systems, it is prone to false positives, misclassifying transitional features from detached or contact binaries as semidetached.

\begin{table}
\caption{Per-class classification metrics (F1 score, Precision, Recall) for morphology classification using XGB. CV metrics are computed on the 845-system cross-validation set (summed across 5 folds). Held-out metrics are computed on the held-out test set; these should be interpreted with caution given the small per-class sample sizes.}
\label{tab:classification}
\begin{tabular}{l ccc c ccc c}
\hline
& \multicolumn{3}{c}{Cross-validation (N=845)} & & \multicolumn{3}{c}{Held-out (N=150)} & \\
\cline{2-4} \cline{6-8}
Morphology & F1 & Precision & Recall & N & F1 & Precision & Recall & N \\
\hline
Detached      & 0.972 & 0.986 & 0.958 & 357 & 0.934 & 0.983 & 0.891 & 64 \\
Semidetached  & 0.916 & 0.899 & 0.935 & 199 & 0.827 & 0.775 & 0.886 & 35 \\
Contact       & 0.959 & 0.955 & 0.962 & 289 & 0.932 & 0.923 & 0.941 & 51 \\
\hline
Overall (macro) & 0.949 & & & 845 & 0.898 & & & 150 \\
\hline
\end{tabular}
\end{table}

\begin{figure}[ht!]
\centering
\plotone{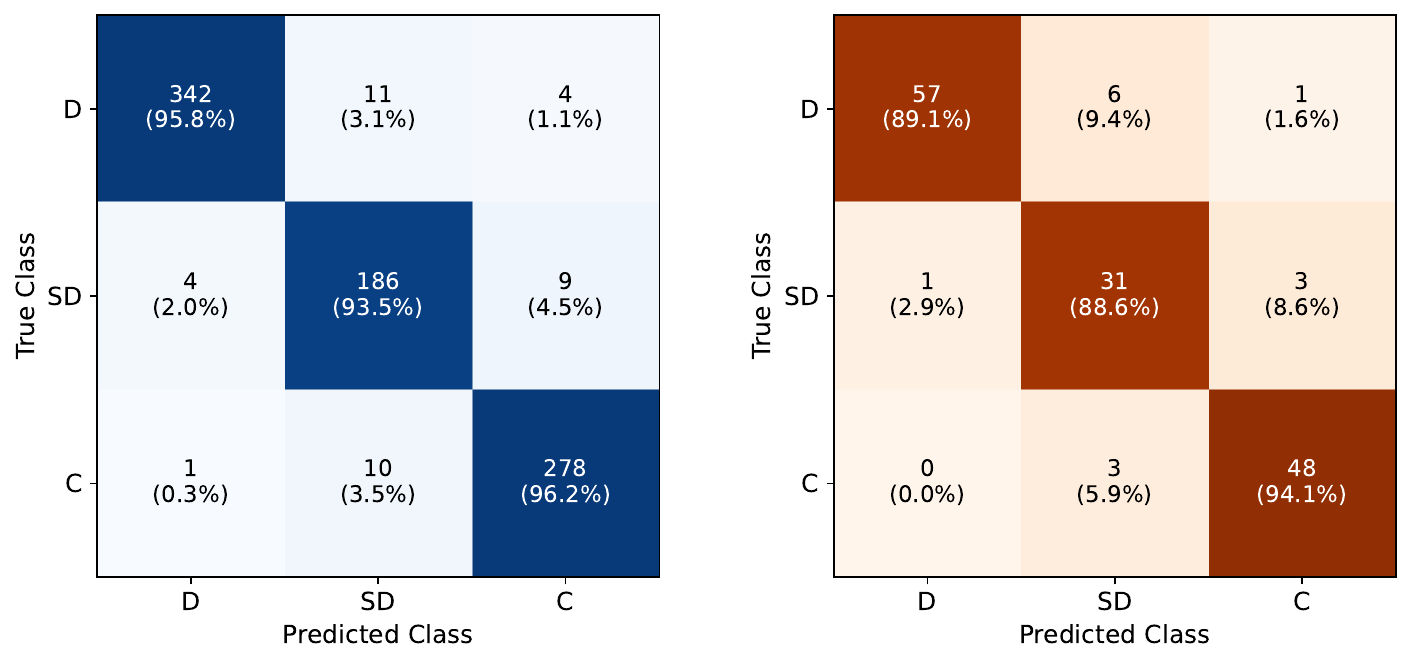}
\caption{Confusion matrices for XGB morphology classification. The left panel shows results on the 5-fold cross-validation set (N = 845, accuracy = 95.4\%), while the right panel shows results on the held-out test set (accuracy = 90.7\%). Each cell gives the number of systems and the row percentage. D, SD and C denote detached, semidetached and contact.}
\label{fig:confusion_matrix}
\end{figure}

Fig.~\ref{fig:confusion_matrix} displays the overall confusion matrix summed across all 5 folds (left panel) and the results on the held-out set (right panel). Of the 845 cross-validation classifications, 806 are correct (95.4\%). The primary confusion occurs between semidetached and detached systems (4 systems misclassified as detached, 11 misclassified from detached to semidetached), accounting for 1.8\% of classifications. A secondary confusion exists between semidetached and contact systems (9 systems misclassified as contact, 10 misclassified from contact to semidetached), representing 2.2\% of cases. Critically, very few systems are misclassified between the morphological extremes (only 1 contact system misclassified as detached and 4 detached systems as contact), confirming that the feature space effectively separates these distinct morphologies. On the held-out set, the classifier achieves 90.7\% accuracy (136/150 correct), with the same confusion pattern: semidetached systems account for the majority of misclassifications (6 detached misclassified as semidetached, 3 contact as semidetached), while direct detached--contact confusion remains minimal (1 detached as contact). The confusion pattern in both sets validates physical expectations: semidetached systems represent an evolutionary transition state, and their light curves can exhibit characteristics intermediate between fully detached and contact configurations.

\begin{figure*}[ht!]
\centering
\plotone{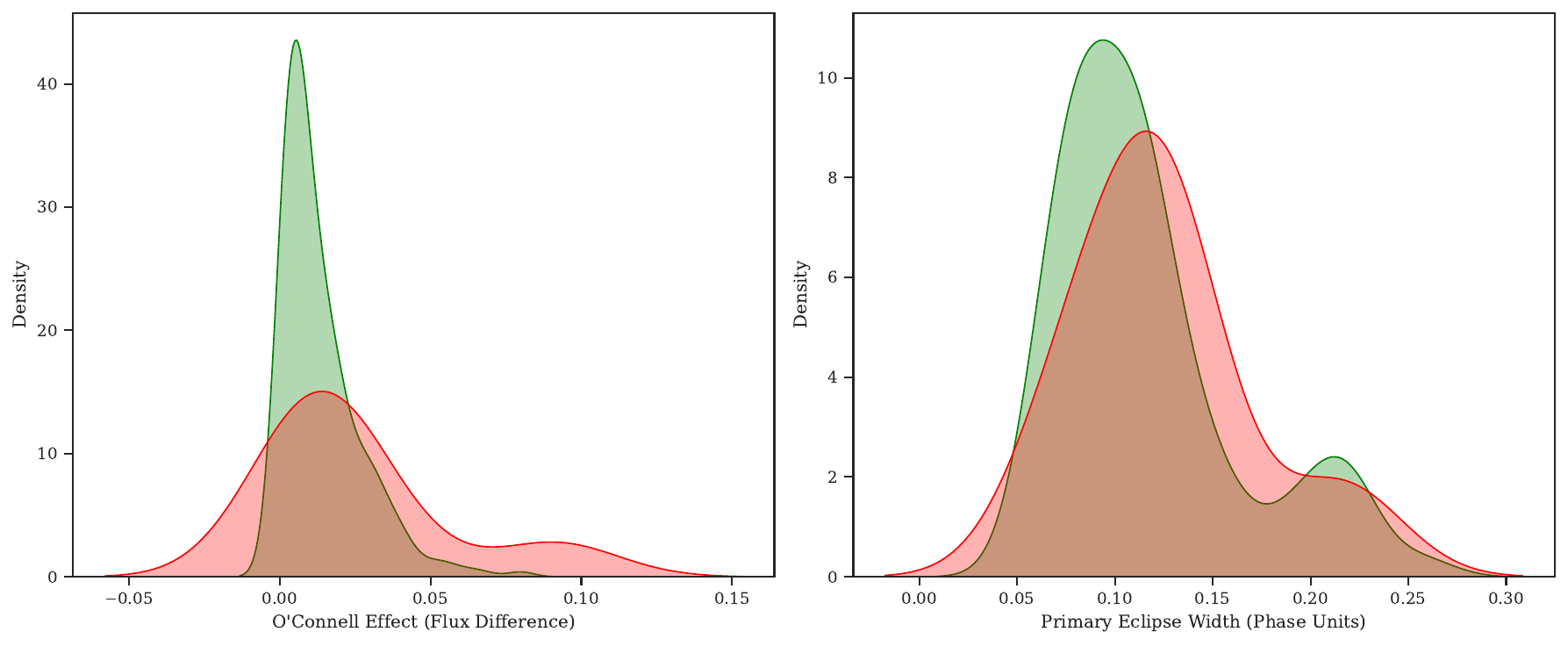}
\caption{Feature distributions for correctly classified (green) and misclassified (red) semidetached systems. Left: O'Connell effect (flux difference between maxima). Right: primary eclipse width (phase units). The y-axis shows normalised probability density.}
\label{fig:sd_error_analysis}
\end{figure*}

The semidetached classification challenge arises from both physical and statistical factors. Physically, semidetached binaries have one star filling its Roche lobe while the other remains detached, creating light curves with features from both regimes: relatively sharp eclipses like detached systems but with mass transfer effects and partial contact signatures. Observationally, these systems span a continuum from nearly-detached to nearly-contact configurations, with no sharp boundary in parameter space. This physical continuity is directly reflected in our model's performance. A post-hoc analysis of the misclassified semidetached systems indicates a strong correlation with the fill-out factor, $f=(\Omega_{in}-\Omega)/(\Omega_{in}-\Omega_{out}$), where $\Omega_{in,out}$ are the critical potentials of the inner and outer Lagrangian points, respectively. The systems misclassified as contact binaries are predominantly 'near-contact' configurations where the secondary component nearly fills the outer Roche lobe ($f\approx0$), producing light curves with continuous variations that mimic overcontact systems. Conversely, the rare semidetached systems misclassified as detached are expected to occupy 'near-detached' configurations ($f < 0$),  where light curves become morphologically indistinguishable from detached binaries without spectroscopic constraints. Our error analysis (Fig.~\ref{fig:sd_error_analysis}) confirms that the misclassified semidetached systems, which  predominantly near-contact configurations,  exhibit broader O'Connell effect distributions and wider primary eclipses compared to correctly classified systems, consistent with their transitional light curve morphology. Statistically, the semidetached class comprises only 23.6\% of the training set (199 systems), providing fewer examples for the models to learn distinguishing patterns compared to the larger detached (357) and contact (289) classes. Despite these challenges, the F1 score of 0.916 indicates that the combination of eclipse features (depths, widths, asymmetries) and Fourier components (ellipsoidal variation amplitudes) successfully captures the morphological signatures across all three classes with minimal systematic confusion.

\section{Predictions on Uncharacterized Systems}
\label{sec:predictions}

We applied our trained XGB framework to two extensive catalogs of eclipsing binary light curves: the OGLE Eclipsing Binary Star database \citep{paw16,sos16} and the {\it{Kepler}} Eclipsing Binary catalogs \citep{prs11,sla12,kir16} database. To handle the computational volume, we developed a GPU-accelerated inference pipeline\footref{fn:github} capable of processing large batches of light curves efficiently.

\subsection{OGLE Predictions}
\label{sec:ogle_prediction}
The pipeline first standardized OGLE light curves designated with the 'ECL' identifier. For systems exceeding 1000 data points, we downsampled the photometry using median binning into 1000 equidistant phase bins. This approach was chosen to preserve the underlying signal structure while effectively suppressing photometric outliers and cosmic ray artifacts. For sparse light curves (fewer than 1000 points), we employed linear interpolation to fill gaps before smoothing the signal with a Savitzky-Golay filter \citep{sav64}. This sequence ensures that high-frequency noise is suppressed without attenuating the eclipse depths, which are critical for temperature ratio estimation. Finally, all light curves were resampled to a uniform 1000-point grid using PCHIP interpolation. We selected this interpolation method specifically to enforce monotonicity, thereby preventing the spurious oscillations (overshoot) that standard cubic splines often introduce near sharp eclipse transitions. Following this preprocessing, feature extraction proceeded identically to the training phase, generating the full set of 51 descriptors for each target. 

A sensitivity analysis was conducted on a subset of the random validation data (N=200) to assess potential systematic bias introduced by switching from the absolute maximum normalization (used in training) to the 99.5th-percentile normalization. We injected synthetic positive outliers into the clean training light curves to simulate the artifacts prevalent in survey data, and then applied the robust percentile normalization. The model predictions for the outlier-injected data were compared against the baseline predictions. The sensitivity analysis confirmed that the switch to robust normalization introduced negligible systematic offsets ($<$1\%). Future scaling of this analysis to the full validation set could further refine these error bounds.

We enforced strict physics-based constraints during the prediction process to ensure astrophysical validity. Specifically, we calculated the inner Lagrangian potential ($\Omega_{\text{in}}$) based on the predicted mass ratio. For systems classified as contact binaries, we imposed the common envelope condition and te potentials  $\Omega_1$ and $\Omega_2$ were set to $\Omega_{\text{in}}$, effectively modeling them as marginal contact systems and eliminating the degeneracy associated with the fill-out factor. For semidetached systems, the secondary potential $\Omega_2$ was fixed to $\Omega_{\text{in}}$, while $\Omega_1$ was constrained to ensure the primary component remained within its Roche lobe.

Confidence scores were assigned to each prediction to quantify reliability. For continuous parameters ($i, q, T_{e_2}/T_{e_1}, \Omega_{1,2}$), the confidence was defined as $1 - \text{CV}$, where CV is the coefficient of variation (see Appendix~\ref{app:formulae}) across the ensemble's five fold-predictions. An overall APC score was derived by averaging the confidences of the geometric parameters, explicitly excluding the morphology classification confidence to prioritize physical robustness. We adopted a confidence threshold of 0.8 based on a preliminary analysis of the validation set, where predictions exceeding this score consistently maintained geometric errors within $\pm$5\%. As a complement to the confidence score, we computed the Mahalanobis distance \citep[$D_\mathrm{M}$,][]{mah36} of each target's feature vector from the training-set centroid, providing an independent measure of how well each system is represented by the training data. Because nearly all OGLE predictions exceed an APC threshold of 0.8 (only two of 104697 fall marginally below), the confidence score alone cannot distinguish reliability tiers; $D_\mathrm{M}$ captures feature-space proximity and is reported alongside each prediction.

\begin{figure*}[ht!]
\centering
\plotone{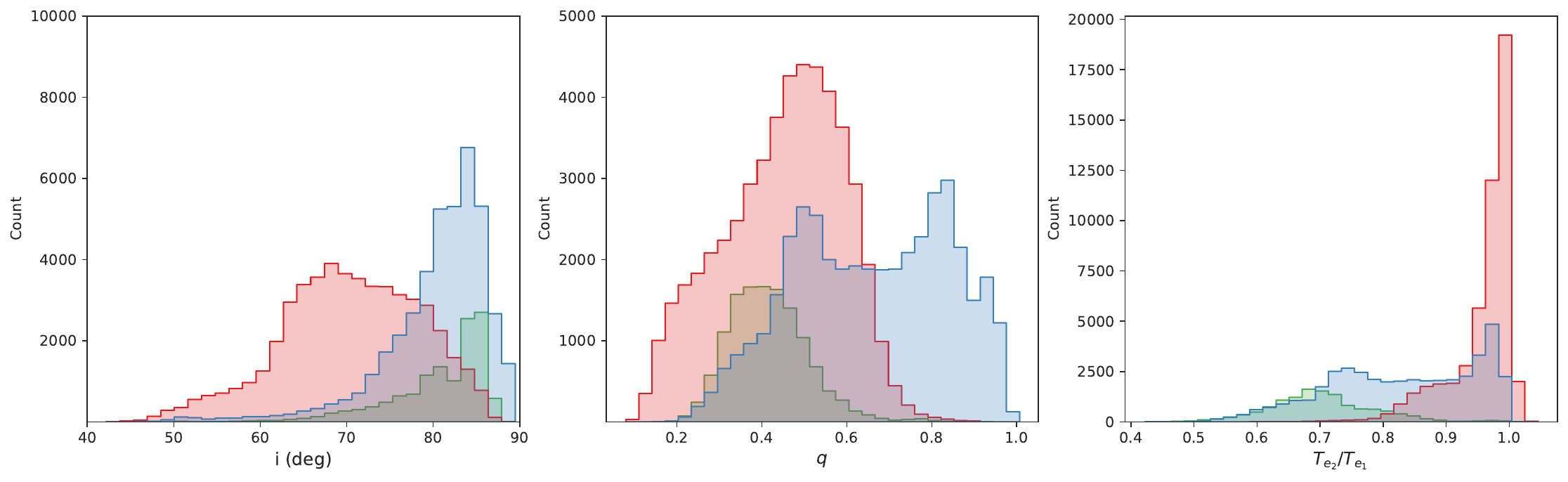}
\caption{Parameter distributions for OGLE predictions. The histograms show the distribution of inclination ($i$), mass ratio ($q$), and effective temperature ratio ($T_{e_2}/T_{e_1}$) separated by morphological class (blue: detached, green: semidetached, red: contact).}
\label{fig:ogle_stats}
\end{figure*}

\begin{figure*}[ht!]
\centering
\plotone{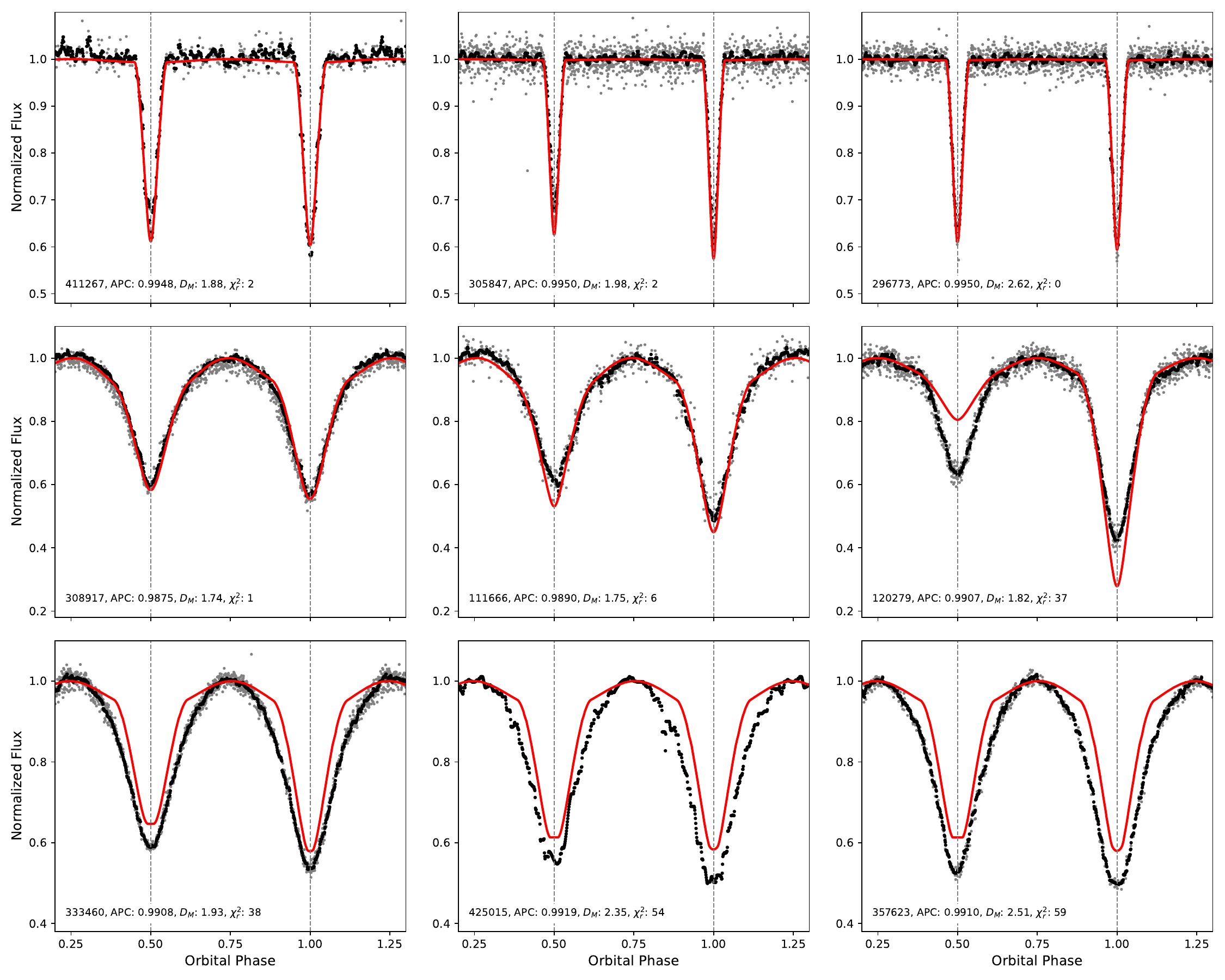}
\caption{Representative OGLE light curves for detached (top), semidetached (middle), and and contact (bottom) systems selected by Mahalanobis distance filtering ($D_\mathrm{M} \leq$ median) and highest APC scores.  Black points are processed light curves; grey dots are the raw catalog data. Vertical dashed lines mark primary and secondary eclipse phases. Panel legends show the OGLE-BLG-ELC identifier, APC score, Mahalanobis distance and reduced $\chi^2_r$. Red lines are synthetic light curves generated directly from the ML-predicted parameters without additional fitting.} 
\label{fig:ogle_grid}
\end{figure*}

The complete catalog of predictions for the OGLE dataset is listed in Table \ref{tab:ogle_predictions}. The model processed a total of 104697 systems. We observed that the entire processed catalog maintained high stability, with candidates consistently. Consequently, our statistical analysis incorporates the full set of predictions, which is characterized by a mean confidence of $0.949 \pm 0.022$. For this population, we find a mean mass ratio of $q = 0.528 \pm 0.186$, a mean inclination of $i = 75.6^\circ \pm 8.6^\circ$, and a mean effective temperature ratio of $T_{e_2}/T_{e_1} = 0.877 \pm 0.123$. The morphological classification identified 41547 detached, 50462 contact, and 12688 semidetached systems. The distributions of these parameters for the high-confidence sample are visualized in Fig.~\ref{fig:ogle_stats}. Representative light curves for the highest-confidence targets in each class are shown in Fig.~\ref{fig:ogle_grid}. When generating these synthetic curves using the Wilson-Devinney based LC code \citep{wil71,wil20}, the effective temperature of the primary component was estimated from the cataloged $(V-I)$ color indices, which we first dereddened using the extinction corrections of \citet{Nat13} to obtain intrinsic values. We utilized the color-temperature calibration of \citet{ram05}. The limb darkening coefficients were determined using the tables of \citet{cla11} whereas the albedos were adopted from \citet{ruc69} and gravity darkening coefficients were derived from \citet{zei24} and Lucy \citet{luc67}, given that the granulation boundary for main-sequence stars is located at about F0 spectral type \citep{gra89}.  Although the observational data were initially normalized, a final multiplicative scaling factor was applied to align the maximum flux level with that of the synthetic light curves. This method ensures a precise comparison of the model to the data while strictly preserving the physics of the eclipse depths. Furthermore, the observational data were slightly phase-shifted to align the primary minimum with phase 1.0, compensating for small uncertainties in the times of minimum and ensuring a strictly geometric comparison of the eclipse morphology. It is important to emphasize that these synthetic light curves represent in the direct application of the model-predicted parameters; no subsequent model refinement or traditional fitting/analysis procedures were employed. We note that the absolute reduced $\chi^2$ values presented in Fig.~\ref{fig:ogle_grid} typically exceed the formal expectation of unity. This is a ubiquitous characteristic of automated modeling applied to high-precision survey data, where formal photometric uncertainties often underestimate systematic noise floors. Furthermore, the synthetic light curves generated for validation assume a static Roche-geometry that intentionally focuses on the dominant geometric signal, ignoring transient second-order variability such as starspots or intrinsic pulsations. Consequently, these $\chi^2$ values should be interpreted as relative metrics for ranking solution fidelity and filtering outliers, rather than as strict probabilistic rejection thresholds. We note that even among the highest-APC systems (Figs.~\ref{fig:ogle_grid}), the reduced $\chi^2$ can reach high order, particularly for contact and semidetached configurations. This reflects the combined effect of the assumed uniform $\sigma = 0.01$ uncertainty, astrophysical variability not captured by the static Roche model, and the photometric degeneracy inherent to these morphologies.

\begin{table*}[ht!]
\centering
\caption{Predictions for 104697 OGLE Eclipsing Binaries. This table lists the classification and geometric parameters estimated by the XGB framework. $C_{m}$ denotes the confidence of the morphological classification. The average parameter confidence (APC) is the mean of the geometric confidences ($C_i, C_q$, etc.), excluding $C_{m}$. $D_\mathrm{M}$ refers to Mahalanobis distance. D, SD and C refers detached, semidetached and contact configurations. Systems with predicted inclinations $i<$70 deg are subject to higher geometric degeneracy inherent to the photometric method. We recommend treating the parameters for these candidates with caution. The metrics are described in Appendix~\ref{app:formulae}. (This table is available in its entirety in machine-readable form in the online journal.)}
\label{tab:ogle_predictions}
\setlength{\tabcolsep}{3pt} 
\begin{tabular}{lcccccccccccccc}
\hline
\hline
OGLE ID & Morph & $C_{m}$ & APC & $D_\mathrm{M}$ &$i$ ($^\circ$) & $C_i$ &  $q$ & $C_q$ &$T_{e_2}/T_{e_1}$ & $C_{T_{e_2}/T_{e_1}}$ & $\Omega_1$ & $C_{\Omega_1}$ & $\Omega_2$ & $C_{\Omega_2}$ \\ 
\hline
OGLE-BLG-ECL-000002&C&1.0&0.941&1.98&75.6&0.99&0.414&0.83&0.984&0.99&2.95&0.95&2.95&0.94\\
OGLE-BLG-ECL-000005&C&1.0&0.981&1.75&83.6&0.99&0.601&0.95&0.909&0.99&3.39&0.99&3.39&0.98\\
OGLE-BLG-ECL-000010&SD&0.8&0.959&2.93&81.1&0.99&0.353&0.95&0.451&0.96&5.97&0.96&2.79&0.94\\
OGLE-BLG-ECL-000012&C&1.0&0.977&1.82&85.3&1.00&0.445&0.94&0.952&1.00&3.02&0.97&3.02&0.98\\
OGLE-BLG-ECL-000013&C&1.0&0.948&3.42&67.1&0.97&0.412&0.90&0.974&0.99&2.94&0.94&2.94&0.94\\
... & ... & ... & ... & ... & ... & ... & ... & ... & ... & ... & ... & ... & ... \\
\hline
\end{tabular}
\end{table*}

An independent validation of the morphological classifier was obtained by cross-matching our OGLE predictions with the OGLE Online Catalog of Variable Stars (OCVS) eclipsing binary classifications \citep{sos16}, which labels systems as contact (C, EW-type) or non-contact (NC, EA/EB-type). Of the $\sim$472000 OCVS entries, 104692 matched our predicted sample. Our classifier correctly identifies 99\% of OCVS-contact systems (recall $= 0.99$), with an overall contact/non-contact accuracy of 71.0\% (F1 macro $= 0.68$). The partial disagreement on non-contact systems (recall $= 0.64$) reflects a fundamental difference in classification criteria: the OCVS assigns NC based on phenomenological light-curve type (EA/EB, i.e.\ distinct eclipse morphology), whereas our model uses Roche-lobe geometry and identifies near-contact configurations as contact even when distinct eclipses remain visible. This primarily affects EB-type (Beta Lyrae) systems, which occupy the transition region between the two classification schemes. Confidence-threshold analysis confirms the disagreements are systematic rather than uncertain: accuracy remains at $\approx$71\% at all morphology classification confidence levels, indicating that our Roche-lobe based criterion and the OCVS phenomenological criterion differ consistently in the EB transition region rather than near a decision boundary.

\subsection{Kepler Predictions}
\label{sec:kepler_prediction}
The framework was subsequently adapted for the {\it{Kepler}} Eclipsing Binary dataset in oreder to assess its performance on space-based observations. The light curves were retrieved from the {\it{Kepler}} Eclipsing Binary Catalog website\footnote{\url{keplerebs.villanova.edu}}, from which we extracted the detrended flux data for our analysis. Consistent with the OGLE pipeline, dense light curves were median-binned and PCHIP interpolation was employed to generate the final standardized input vector, ensuring robust handling of the irregular sampling inherent in {\it{Kepler}} data.

\begin{figure*}[ht!]
\centering
\plotone{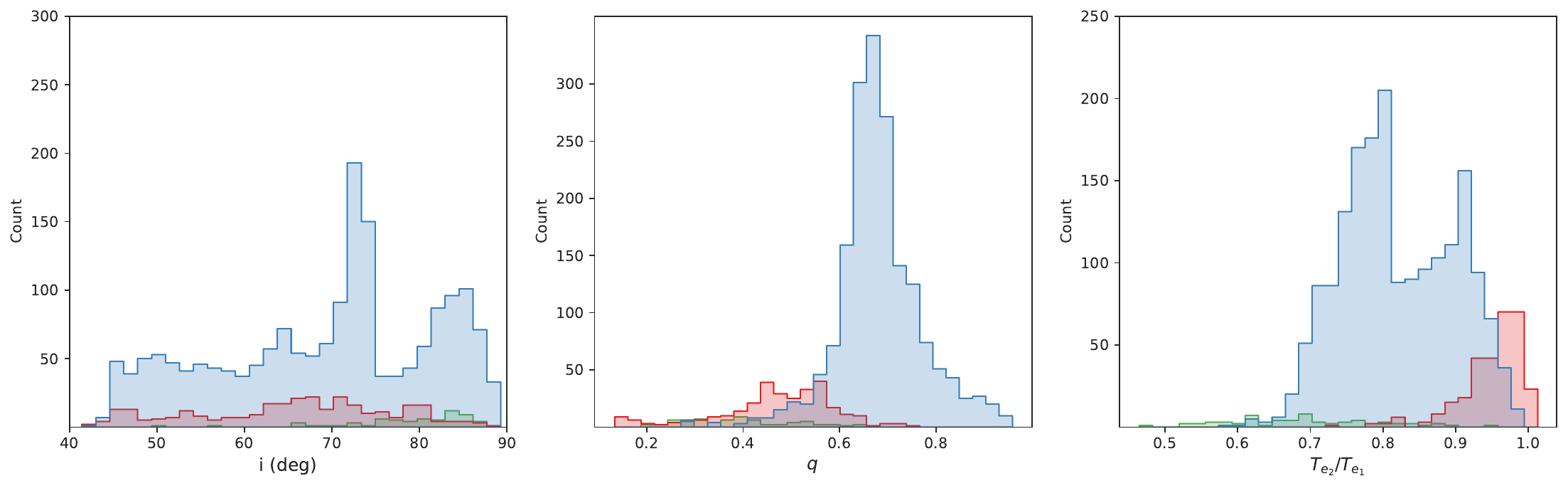}
\caption{Same as Fig.~\ref{fig:ogle_stats}, but for 2157 {\it{Kepler}} eclipsing binary systems.}
\label{fig:kepler_stats}
\end{figure*}

\begin{figure*}[ht!]
\centering
\plotone{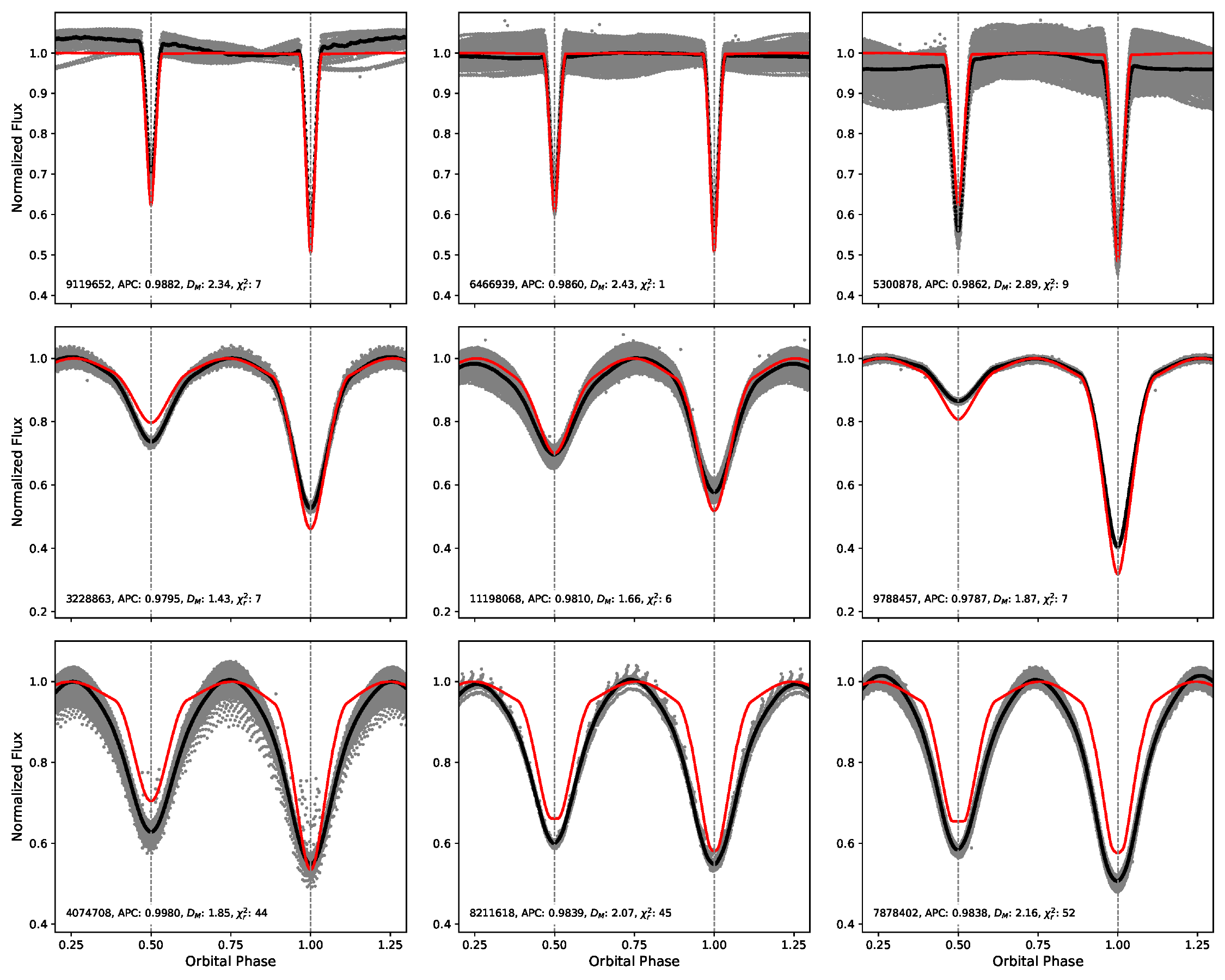}
\caption{Same as Fig.~\ref{fig:ogle_grid}, but for high-confidence predictions from the {\it{Kepler}} eclipsing binary dataset.}
\label{fig:kepler_grid}
\end{figure*}

The same physics constraints and confidence scoring metrics applied to the OGLE data were used for {\it{Kepler}} targets. The model successfully recovered parameters for detached, semidetached, and contact systems. The complete set of {\it{Kepler}} predictions is provided in Table \ref{tab:kepler_predictions}. The model processed a total of 2157 \textit{Kepler} eclipsing binary systems.  All predictions exceed the APC $> 0.8$ threshold, with an average parameter confidence of $0.913 \pm 0.034$. We derive a mean mass ratio of $q = 0.643 \pm 0.127$, an inclination of $i = 69.0\degr \pm 12.1\degr$, and an effective temperature ratio of $T_{e_2}/T_{e_1} = 0.835 \pm 0.090$. Morphologically, the sample is dominated by detached systems ($N = 1792$), followed by contact ($N = 302$) and semidetached ($N = 63$) binaries. As in the OGLE analysis (Sec.~\ref{sec:ogle_prediction}), the APC alone cannot distinguish reliability tiers since all 2157 predictions exceed~0.8; we therefore complement it with the Mahalanobis distance ($D_\mathrm{M}$), reported alongside each prediction in Table~\ref{tab:kepler_predictions}. The statistical distributions for these parameters are presented in Fig.~\ref{fig:kepler_stats} and the nine representative systems shown in Fig.~\ref{fig:kepler_grid} were selected by filtering to $D_\mathrm{M}$  below the median and then ranking by APC within each morphological class. To generate the light curves using predicted parameters via LC code, we followed the same procedure used for the OGLE data, with the exception that the effective temperatures for the {\it{Kepler}} targets were adopted directly from the {\it{Kepler}} Eclipsing Binary catalog to ensure consistency. 

\begin{table*}[h!]
\centering
\caption{Predictions for 2157 {\it{Kepler}} Eclipsing Binaries. Columns follow the same definitions as Table \ref{tab:ogle_predictions}. (This table is available in its entirety in machine-readable form in the online journal.)}
\label{tab:kepler_predictions}
\setlength{\tabcolsep}{3pt} 
\begin{tabular}{lcccccccccccccc}
\hline \hline
{\it{Kepler}} ID & Morph & $C_{m}$ & APC & $D_\mathrm{M}$ &$i$ ($^\circ$) & $C_i$ &  $q$ & $C_q$ &$T_{e_2}/T_{e_1}$ & $C_{T_{e_2}/T_{e_1}}$ & $\Omega_1$ & $C_{\Omega_1}$ & $\Omega_2$ & $C_{\Omega_2}$ \\
\hline
1026032&D&1.0&0.895&6.05&77.3&0.93&0.704&0.91&0.760&0.98&22.72&0.85&24.39&0.81\\
1026957&D&1.0&0.892&10.75&54.0&0.95&0.713&0.95&0.796&0.96&5.97&0.92&12.46&0.69\\
1295531&D&0.8&0.887&16.58&57.7&0.91&0.636&0.90&0.921&0.99&5.77&0.95&12.88&0.70\\
1432214&D&1.0&0.898&10.33&57.0&0.92&0.643&0.93&0.800&0.95&6.08&0.92&6.98&0.77\\
1433410&C&0.6&0.974&3.86&70.3&0.98&0.633&0.96&0.973&0.99&3.46&0.97&3.46&0.97\\
... & ... & ... & ... & ... & ... & ... & ... & ... & ... & ... & ... & ... & ... \\
\hline
\end{tabular}
\end{table*}

We observe that the estimation accuracy varies significantly by morphological type, with contact binaries (bottom panels of Fig.~\ref{fig:kepler_grid}) showing lower coherence than detached systems. This discrepancy is primarily driven by the inherent physical degeneracies of contact configurations. Unlike detached systems, where distinct ingress and egress discontinuities tightly constrain the geometric parameters, contact binaries exhibit almost continuous light variation. This lack of sharp geometric features introduces a significant degeneracy between the mass ratio and inclination, where multiple parameter combinations can reproduce nearly identical photometric signatures. Consequently, the machine learning model struggles to distinguish the unique physical solution within this ambiguous parameter space. This physical ambiguity is further compounded by the necessary smoothing of the sharp, non-linear minima during preprocessing, which can subtly attenuate the amplitude constraints. Notably, the APC and $\chi^2$ metrics are not always correlated for contact systems as this parameter-space ambiguity allows models to disagree on parameters (lowering APC) while each degenerate solution reproduces the light curve comparably well, and vice versa. The resulting model predictions occasionally deviate from the data because the model's predicted parameters represent a physically degenerate solution that fits the global morphology but misses these subtle amplitude constraints. Additionally, the elevated reduced $\chi^2$ values are an expected consequence of the exceptional photometric precision of the Kepler data, where even minute deviations are statistically resolved.

Cross-matching our predictions with the machine learning based parameters of the {\it{Kepler}} Eclipsing Binary Catalog \citep{prs11} allowed us to evaluate the model's generalization capability across different instruments. For the 980 overlapping systems, our pipeline achieved a morphology classification accuracy of 77\%, with strong performance for detached systems (recall~0.97) and high precision for contact identification (precision~0.96); semidetached recovery remains limited (recall~0.17, $N=83$), reflecting both the small catalog sample and the morphological ambiguity of this transitional class. In the continuous parameter space, we observe systematic deviations, particularly in the mass ratio (available only for contact binaries, MAE$=$0.52) and effective temperature ratio (MAE$=$0.11). These discrepancies are attributable to three fundamental factors. First, the catalog values used for comparison are themselves predictions from a neural network trained on synthetic light curves, whereas our model is trained on a diverse compilation of observational light curves. Consequently, our model implicitly accounts for complex astrophysical noise such as spot modulation and third-light contamination which is absent in idealized synthetic training sets. Second, \citet{prs11} explicitly acknowledge that their photometric mass ratios suffer from 'strong signs of systematics' and are largely unconstrained for partially eclipsing systems, noting this as an 'inherent limitation' of their photometric method. Our model's tendency to cluster contact binary mass ratios near $q\approx$ 0.5 likely reflects a conservative prior learned from real populations, avoiding the overfitting of degenerate light curve features often seen in synthetic-only models. Finally, the systematic offset in $T_{e_2}/T_{e_1}$ highlights the physical domain shift between our training compilations and the {\it{Kepler}} bandpass. Since effective temperature ratios determine eclipse depths in a wavelength-dependent manner, a direct comparison between parameters derived from heterogeneous photometric systems will naturally exhibit systematic offsets.

We performed an additional cross-match with the catalog of \citet{win19}, who derived orbital and stellar parameters for 728 detached and semidetached Kepler eclipsing binaries through joint Bayesian modeling of curves and spectral energy distributions using PARSEC stellar isochrones. After selecting systems with $q < 1.05$ to remain consistent with our training range, we cross-matched 499 systems. For the mass ratio, our model achieves MAE $= 0.156$ and Pearson correlation coefficient $r = 0.473$, measuring the linear agreement between predicted and catalog values, with 71.3\% of predictions within 0.2 of the catalog value. For the temperature ratio, we obtain MAE $= 0.101$ and $r = 0.694$, with 60.5\% of predictions within 0.10, consistent with the Pr\v{s}a et al. comparison. For inclination, 14.6\% of catalog systems report $i > 90\degr$, values outside our training domain ($50\degr$--$90\degr$). Even after folding these to $180\degr - i$, a systematic offset of approximately $-5\degr$ persists (MAE $= 6.6\degr$), with our model consistently underestimating the inclination. This negative bias is consistent with the expected effect of unmodeled third-light dilution discussed in Section 5, which reduces observed eclipse depths and forces lower inclination estimates.

\section{Discussion and Conclusion}
\label{sec:discussion}

In this study, we addressed the growing disparity between the accumulation of photometric survey data and the capacity for detailed analysis by developing an integrated machine learning framework. By training ensemble models on a curated set of well-characterized systems, we demonstrated that domain-specific feature engineering allows for the simultaneous classification of morphology and estimation of physical parameters effectively. The generally good correlation between predicted and true values, particularly for the effective temperature ratio and surface potentials, suggests that the information content of a single-band light curve is sufficient to constrain the underlying geometry of eclipsing binaries.

\begin{figure*}[ht!]
\centering
\epsscale{0.7}
\plotone{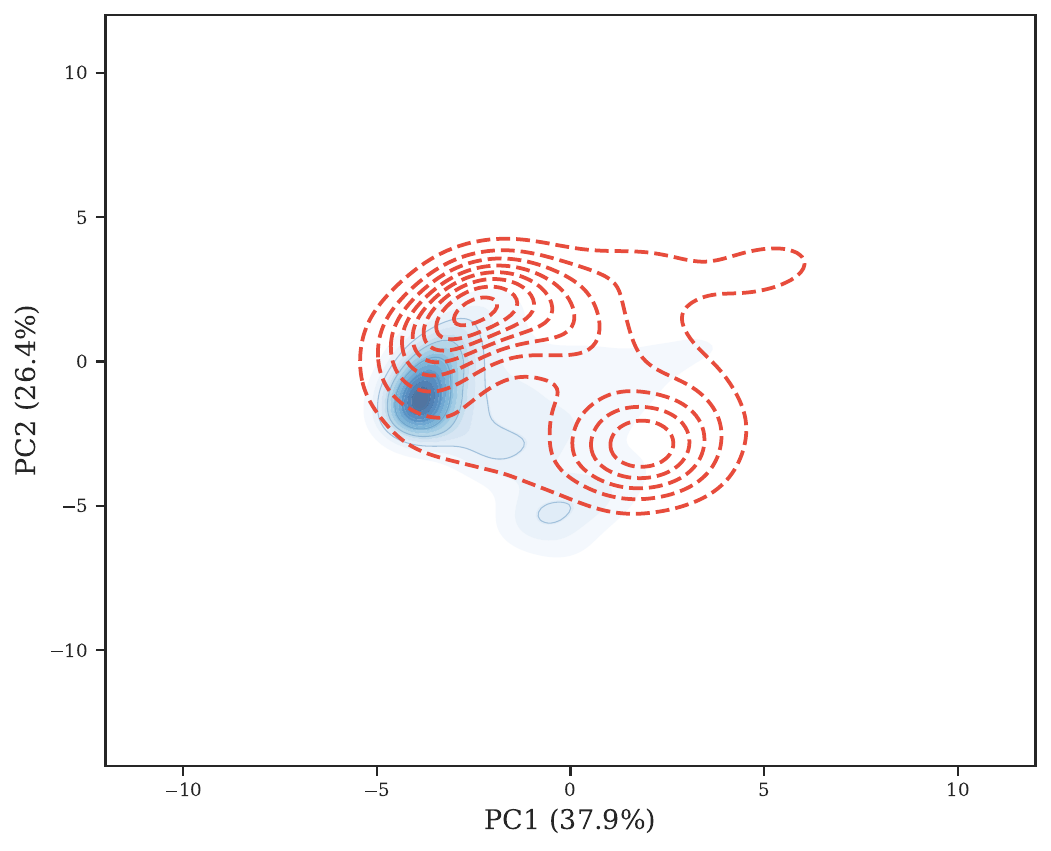}
\caption{Parameter-space coverage analysis in PCA space (PC1 + PC2 account for 64.3\% of variance). The filled blue density shows the KDE of 10000 randomly sampled OGLE survey targets; the red dashed contours represent the KDE of the 995 training systems. The two-dimensional KDE overlap coefficient is OVL = 0.47.}
\label{fig:pca}
\end{figure*}

A fundamental question in applying supervised learning to large-scale surveys is whether the training sample, 995 well-characterized systems in our case, is sufficiently representative of the vast, unclassified population. We address this question by performing a PCA on the extracted feature vectors and comparing the density distributions of both samples. Fig.~\ref{fig:pca} projects both the training set and a random subsample of 10000 unclassified OGLE light curves onto the first two principal components, which account for 64.3\% of the total variance. A systematic offset is visible, with the training sample concentrating at higher PC1 values relative to the survey core. Quantitatively, pairwise KS tests reject the null hypothesis that both samples are drawn from the same distribution along both PC1 ($D=0.28$, $p<10^{-60}$) and PC2 ($D=0.36$, $p<10^{-100}$), and the two-dimensional KDE overlap coefficient is 0.47. This offset is physically expected since the training set consists of bright, well-characterized systems with deep eclipses and high orbital inclinations (Section~\ref{sec:data}), whereas the OGLE survey spans a much broader range of signal quality and eclipse morphology. Crucially, despite this distributional difference in feature space, the model generalizes effectively to the survey population, as demonstrated by the held-out validation ($R^2 = 0.77$--$0.92$; Table~\ref{tab:xgboost_regression}), the OGLE morphology cross-match (Sec.~\ref{sec:ogle_prediction}), and the independent Kepler comparison (Sec.~\ref{sec:kepler_prediction}). This is, in fact, a stronger indicator of robustness than distributional overlap alone would provide. It demonstrates that the physical relationships encoded by the model, specifically the mapping from light-curve morphology to Roche geometry, generalize beyond the exact training distribution to regions of feature space not densely sampled during training. Nonetheless, predictions for systems lying far from the training distribution should be treated with greater caution, as reflected in their higher Mahalanobis distances. Both our training sample and the OGLE/\textit{Kepler} targets are subject to the same inclination-selection effect that preferentially detects edge-on systems. Extending the training set to include lower-inclination and fainter systems would be a natural direction for future work.

We also note a limitation of the APC confidence metric as currently defined. Because APC measures the consistency of predictions across the five cross-validation fold models rather than their absolute accuracy, a systematically biased but internally consistent ensemble would still yield high APC values. In practice, we observe that nearly all 104697 OGLE predictions (all but two) and all 2157 Kepler predictions exceed the APC $> 0.8$ threshold, meaning the metric retains 100\% of candidates and provides limited discriminating power for filtering unreliable predictions. This ceiling effect arises because the five fold models, trained on largely overlapping data (80\% shared), naturally converge to similar solutions. To partially address this limitation, we introduced the Mahalanobis distance ($D_\mathrm{M}$, see Sec.~\ref{sec:ogle_prediction} and \ref{sec:kepler_prediction}) as a complementary reliability metric that captures how well each target's feature vector is represented by the training data, independent of model agreement. Future work could improve prediction filtering by incorporating the $\chi^2$ goodness-of-fit as a complementary quality metric.

The application of our framework to the OGLE and {\it{Kepler}} datasets demonstrates the potential of this approach. We successfully recovered physical parameters for over 100000 systems, producing distributions that align with astrophysical expectations. While the model exhibits higher uncertainty for low-inclination systems, which is a known limitation of photometric solutions where eclipses provide weaker geometric constraints, the high-confidence predictions offer a robust starting point for detailed analysis. We also observe a discrepancy in the recovery rate of semidetached systems between the OGLE (12.1\%) and {\it{Kepler}} (2.9\%) catalogs. This lower yield in the {\it{Kepler}} sample may stem from the morphological ambiguity of semidetached configurations, which span a continuous parameter space between detached and contact geometries. The specific smoothing applied to {\it{Kepler}}'s data may subtly attenuate the sharp ingress/egress features required to distinguish 'near-contact' semidetached systems from overcontact binaries. Furthermore, the Kepler Eclipsing Binary catalog's target selection function differs fundamentally from OGLE's unbiased survey approach. As noted by \citet{prs11}, the Kepler target list was pre-selected to prioritize main-sequence stars suitable for planet detection, systematically omitting evolved giants and stars in overcrowded fields. While dynamically evolved contact binaries (W~UMa types) are comprised of main-sequence dwarfs and thus retained, their relative abundance in the \textit{Kepler} catalog ($\approx$14\%) appears lower than in ground-based surveys ($\approx$48\%). This may not be due to exclusion, but rather to a detection bias reversal: Kepler's superior photometric precision reveals a vast population of low-amplitude detached binaries that are invisible to ground-based surveys, effectively diluting the contact binary fraction.

To contextualize our results within the rapidly evolving landscape of automated binary characterization, Table \ref{tab:comparison} compares our framework with recent key studies. The deep learning architectures have achieved strong results in specific tasks such as the $>$96\% classification accuracy for detached and contact systems reported by \citet{Parimucha2024} using CNNs, or the high-fidelity light curve reconstruction for contact binaries achieved by \citet{Ding2024}. Most of existing pipelines are specialized for either morphological classification or single-class parameter estimation. In contrast, our XGB ensemble offers a holistic solution. It maintains competitive classification accuracy (95.4\%) comparable to the deep learning benchmarks while simultaneously solving the inverse problem for physical parameters across detached, semidetached, and contact configurations. Although specialized neural networks like those of \citet{Li2025} may offer higher sensitivity for specific phenomena like the O'Connell effect in contact binaries, our feature-based approach provides a balanced trade-off, delivering robust, physically interpretable estimates ($R^2$=0.77-0.92) for the broad diversity of systems found in large-scale surveys like OGLE and {\it{Kepler}}.

\begin{table*}[ht!]
\centering
\caption{Comparison of our work with recent machine learning frameworks for eclipsing binary systems. Unlike studies focused solely on classification or specific binary subclasses, our framework integrates morphology classification and parameter estimation for detached (D), semidetached (SD), and contact (C) systems.}
\label{tab:comparison}
\begin{tabular}{lcccc}
\hline
Study & Methodology & Scope & Task & Metric \\
\hline
\citet{Che20} & DBSCAN & Periodic Vars. & Classification & Acc $\approx 98\%$ \\
\citet{Parimucha2024} & CNN (ResNet50) & D, C & Classification & Acc $> 96\%$ \\
\citet{Shan2025} & Hybrid (CNN+LSTM) & D, SD, C & Detection & Recall=$99.1\%$ \\
\citet{Ding2024} & Autoencoder & Contact & Detection & $R^2_{recon} > 0.99^a$ \\
\citet{Li2025} & Neural Network & Contact & Param. Est. & $R^2 = 0.999^b$ \\
\hline
This study & XGB Ensemble & D, SD, C & Class. + Params & Acc $>$95\%, $\bar{R}^2 > 0.87^c$\\
\hline
\end{tabular}
\par\smallskip
\begin{minipage}{\linewidth}
\footnotesize\raggedright
{\footnotesize
$^a$ The coefficient of determination measuring the Autoencoder model's light curve reconstruction quality.\\
$^b$ The average goodness of fit on synthetic testing datasets.\\
$^c$ The classification accuracy and the overall average coefficient of determination score across all five physical parameters.}
\end{minipage}
\end{table*}

A critical advantage of data-driven approaches is their computational efficiency. Conventional light curve analysis methods, often require significant manual intervention and computation time ranging from minutes to hours per system. In contrast, our GPU-accelerated XGB inference pipeline processed the entire catalog of 104697 systems in approximately 30 minutes, averaging just 17 ms per light curve on an NVIDIA RTX A4000 GPU. This represents a speedup of several orders of magnitude compared to traditional iterative modeling, enabling the rapid and efficient characterization of massive datasets. This speed enables the rapid scanning of massive datasets from past/current and future space missions (e.g., TESS, PLATO) to identify candidates of interest for follow-up study.

While the current results are promising, the framework remains open to refinement, particularly regarding complex astrophysical signals. For instance, our preprocessing pipeline aligns the primary eclipse to phase 1.0; while this standardizes the input, it optimizes the model for circular orbits and may result in reduced accuracy for highly eccentric detached systems where the secondary eclipse is shifted. Similarly, although we utilize O'Connell effect features to capture asymmetries, the model currently maps these to geometric parameters rather than explicit spot models, which can introduce systematic bias in the mass ratio for heavily spotted systems. Furthermore, since our model does not include a parameter for extrinsic flux, any unmodeled light arising from a physical tertiary companion or simply background contamination in crowded fields will dilute the observed eclipse depths. This unmodeled dilution forces the geometric model to compensate by underestimating the orbital inclination. Future iterations of this work will address these regimes by incorporating third light and spot parameters directly into the training set. Additionally, we plan to explicitly analyze residuals as a function of orbital period and, most critically, rigorously validate our photometric mass ratios against independent spectroscopic radial velocity solutions from surveys such as LAMOST. Nevertheless, this work establishes that automated, high-speed characterization of eclipsing binaries is not only feasible but essential for the modern era of time-domain astronomy. 

Finally, returning to the question posed in our title, our results suggest a promising, though qualified, affirmative: a modest but well-characterized training sample demonstrates the potential to bridge the gap between detailed individual modeling and massive survey characterization.

\begin{acknowledgments}
We thank the anonymous referee for their constructive comments, which improved the overall manuscript. This study was partly supported by the Turkish Scientiﬁc and Research Council (T{\"U}B{\.I}TAK~125F087). The numerical calculations reported in this paper were partially performed at T{\"U}B{\.I}TAK ULAKBIM, High Performance and Grid Computing Center (TRUBA resources). The author acknowledges the use of Claude Code (Anthropic, model Claude Sonnet 4.5) and Gemini (Google, model 3 Pro) for technical assistance in writing the code for the data processing pipeline and debugging the machine learning workflows. These tools were not used for generating research data. Crucially, all AI-generated code was manually audited and verified for scientific accuracy by the author to ensure the integrity of the computational results. This research has made use of the VizieR catalogue access tool, CDS, Strasbourg, France. This work is based on observations obtained with the 1.3-m Warsaw telescope at Las Campanas Observatory, Chile, operated by the Carnegie Institution for Science. The OGLE project is supported by the Polish National Science Centre. This paper includes data collected by the {\it{Kepler}} mission. Funding for the {\it{Kepler}} mission is provided by the NASA Science Mission Directorate.
\end{acknowledgments}

\bibliography{Ulas_bib}{}
\bibliographystyle{aasjournalv7}

\appendix

\section{Formulation for Evaluation Metrics}
\label{app:formulae}

To quantify the reliability of our predictions for uncharacterized systems, we formulated confidence scores based on the consensus of the ensemble. For continuous regression parameters ($q, i, T_{e_2}/T_{e_1}, \Omega_1,\Omega_2$), we utilized the predictions from the $K=5$ independent fold models. Let $y_k$ be the prediction of the $k$-th model for a given parameter. We first calculate the ensemble mean ($\mu$) and standard deviation ($\sigma$):

\begin{equation}
    \mu = \frac{1}{K} \sum_{k=1}^{K} y_k, \quad \sigma = \sqrt{\frac{1}{K-1} \sum_{k=1}^{K} (y_k - \mu)^2}
\end{equation}

The Coefficient of Variation ($CV$) represents the dispersion of the predictions relative to their magnitude:

\begin{equation}
    CV = \frac{\sigma}{\mu}
\end{equation}

The parameter-specific confidence score ($C_{\theta}$) is defined as:

\begin{equation}
    C_{\theta} = \max\left(0, 1 - CV\right)
\end{equation}

For the morphological classification, the confidence score $C_{m}$ is derived directly from the softmax probability distribution of the XGB classifier. Let $P(c)$ be the predicted probability for the winning class $c$:

\begin{equation}
    C_{m} = \max_c P(c)
\end{equation}

To select high-quality candidates for the final catalog, we define an Average Parameter Confidence score (APC) as the arithmetic mean of the confidence scores for the five continuous physical parameters, explicitly excluding the morphology classification confidence:

\begin{equation} 
    APC = \frac{1}{5} \sum_{\theta} C_{\theta} 
\end{equation}
where $\theta \in \{i, q, T_{e_2}/T_{e_1}, \Omega_1, \Omega_2\}$.

To evaluate the agreement between the observed light curve and the synthetic model generated from our predicted
parameters, we compute the $\chi^2$ statistic. We adopt a uniform photometric uncertainty of $\sigma = 0.01$ in normalized flux units
for all observations:

\begin{equation}
    \chi^2 = \sum_{j=1}^{N} [(F_{obs,j} - F_{calc,j}) / \sigma]  ^2
\end{equation}
To normalize the goodness-of-fit metric for the number of data points and model complexity, we report the reduced chi-squared statistic:

\begin{equation}
    \chi^2_{r} = \frac{\chi^2}{N - k}
\end{equation}
where $N$ is the number of photometric observations and $k = 5$ represents the five predicted physical parameters.

\end{document}